\documentclass[manuscript]{aastex}
\usepackage{apjfonts}
\usepackage{epsfig}
\usepackage{graphics}
\usepackage{natbib,graphicx}

\newcommand{\rhostar}{\ensuremath{\rho_\star}}

\newcommand{\ik}{{\it Kepler~}}

\newcommand{\logg}{\ensuremath{\log{g}}}

\newcommand{\rstar}{\ensuremath{R_\star}}

\begin{document}
\title{KOI-152's Low Density Planets}

\author{Daniel Jontof-Hutter\altaffilmark{1}, Jack J. Lissauer\altaffilmark{1}, Jason F. Rowe\altaffilmark{1,2}, Daniel C. Fabrycky\altaffilmark{3}}

\email{Daniel.S.Jontof-Hutter@nasa.gov}
\altaffiltext{1}{NASA Ames Research Center, Moffett Field, CA 94035, USA}
\altaffiltext{2}{SETI Institute/NASA Ames Research Center, Moffett Field, CA 94035, USA}
\altaffiltext{3}{Department of Astronomy and Astrophysics, University of Chicago, 5640 South Ellis Avenue, Chicago, IL 60637, USA}

\begin{abstract}
KOI-152 is among the first known systems of multiple transiting planetary candidates \citep{ste10} ranging in size from 3.5 to 7 times the size of the Earth, in a compact configuration with orbital periods near a 1:2:4:6 chain of commensurability, from 13.5 days to 81.1 days. All four planets exhibit transit timing variations with periods that are consistent with the distance of each planet to resonance with its neighbors. We perform a dynamical analysis of the system based on transit timing measurements over 1282 days of \textit{Kepler} photometry. Stellar parameters are obtained with a combination of spectral classification and the stellar density constraints provided by light curve analysis and orbital eccentricity solutions from our dynamical study. Our models provide tight constraints on the masses of all four transiting bodies, demonstrating that they are planets and that they orbit the same star. All four of KOI-152's transiting planets have low densities given their sizes, consistent with other studies of compact multiplanet transiting systems. The largest of the four, KOI-152.01, has the lowest bulk density yet determined amongst sub-Saturn mass planets. 
\end{abstract}

\section{Introduction}
Within our Solar System, Earth and smaller bodies are primarily mixtures of refractories, rock and metals. In the outer solar system, bodies that are too small to retain deep atmospheres contain rock and ices. In the larger planets, including Uranus and Neptune, the light elements H and He dominate the volume. There no local examples of bodies intermediate in size or mass between Earth ($1R_{\oplus}$, $1M_{\oplus}$) and Uranus/Neptune, both of which are larger than 3.8 $R_{\oplus}$ and more massive than 14 M$_{\oplus}$. Mass determinations of transiting exoplanets are beginning to allow the characterization of planets in this size range. As more planetary masses and radii are measured, their bulk densities will provide more constraints on their compositions.

To derive meaningful planetary densities requires both accurate mass and radius determinations. The ratio of the planetary radius to the stellar radius is a direct measurement from transit light curves, where the fraction of starlight blocked during transit is a simple measure of the projected area of the planet. The uncertainty in this measurement typically rests on the accuracy to which the stellar radius can be constrained. The star 55 Cancri is unique as the host of a transiting sub-neptune exoplanet in having a direct measurement of its radius via interferometry \citep{von11}. High resolution spectral classification of the atmosphere of \textit{Kepler} host stars gives a model dependent measurement of the stellar radius with an uncertainty of typically 10\%.


The measurement of stellar radius from spectral classification and modelling can be improved upon with additional information. The gold standard for this purpose is where asteroseismic oscillations are detected in the photometric light curve. Amongst these stars, uncertainties in mass and radius can be reduced to $\sim 1\%$, although detections are only available for giant stars and the brightest dwarf stars \citep{hub13}.

 Another constraint on the stellar density, and hence its radius, can be gleaned from the orbital constraints of exoplanets. The scaled semi-major axis, $a$/\rstar, (where $R_{\star}$ is the stellar radius) can be estimated roughly from the transit and ingress durations, but accurate measurement of $a$/\rstar\ requires additional information about the orbital eccentricity and alignment ($\omega$), the longitude of pericenter from the passage of the orbiting planet through the sky-plane towards the Earth. For a measured fractional transit depth $\delta$, transit duration $T$, and ingress or egress duration $\tau$,
\begin{equation}
\frac{a}{\rstar} = \frac{\delta^{1/4}}{\pi}\frac{P}{\sqrt{T\tau}}\left( \frac{\sqrt{1-e^2}}{1+e\sin\omega}\right),
\label{arstar}
\end{equation}
\citep{winn11}. 
 For the purposes of measuring the stellar radius, information about the the orbital eccentricity from dynamical fits, used in Eq.~\ref{arstar}, can provide an independent constraint on the stellar bulk density following Kepler's Third Law (\citealt{sea03,winn11}):
\begin{equation} 
\rhostar \approx \frac{3\pi}{G P^2} \left( \frac{a}{\rstar} \right)^3.
\label{rhostar}
\end{equation}
Here $\rhostar$ is the bulk density of the star, $G$ is the gravitational constant, and $P$ is the orbital period of the transiting planet. In their study of Kepler-11, \citet{liss13} used dynamical solutions for the orbital eccentricities, alongside high resolution spectra to reduce the uncertainty on the stellar radius to 2\%. The planetary radii were measured with nearly that precision.

Of the transiting exoplanets, the majority of mass determinations to date are the result of radial velocity (RV) spectroscopy, and amongst these, most are the so-called hot jupiters, planets with substantial envelopes that orbit very close to their host star. The transiting neptunes and sub-neptunes with measured RV masses all have short orbital periods, the longest being Kepler-68 b, \citep{gil13} at 5.4 days. Note that Kepler-18's planets have had their masses measured by the combined constraints of RV and transit timing variations (TTVs), the farthest one orbiting every 14.9 days \citep{coch11}.

TTVs exploit the high degree of accuracy in measuring the transit times of transiting exoplanets, with transit time uncertainties as low as a few minutes in some cases. These probe interplanetary perturbations, and in general are sensitive to the mass ratio of perturbing neighboring planets to the host star. The strongest signals in TTVs occur when planets are near (but not trapped in) mean motion resonances, and the resonant argument cycles with a periodicity that is well sampled over a baseline of transit timing measurements. Near first order resonances, the coherence time is long enough for perturbations to build constructively to an easily detectable amplitude. Too far from resonance, the perturbations lose their coherence rapidly, and the TTV amplitude is reduced. Too near resonance, \textit{Kepler}'s four years of observations do not cover a complete cycle. Nevertheless, since TTVs are sensitive to interplanetary perturbations and not the effect of the planet on the star, planetary masses can be determined to far greater orbital periods than RV, as long as enough transit times have been measured to detect timing variations. The super-neptune Kepler-30 d, with an orbit period of 143 days \citep{san12} has the longest orbital period for an exoplanet with a mass determination on the mass-radius diagram. The difference in orbital periods probes by these two separate techniques highlight their respective biases. RV planets orbit close to their star, are hotter, and several, particularly the earths; including Kepler-10 b \citep{bat11}, CoRoT-7 b \citep{fer11}, KOI-94 b \citep{weis13}, Kepler-20 b \citep{gau12} and potentially Kepler-18 b \citep{coch11} seem to lack deep atmospheres. The TTV mass determinations, such as Kepler-11 (\citealt{liss11a,liss13}), Kepler-18 \citep{coch11}, Kepler-30 \citep{san12} and Kepler-36 \citep{car12} are all either compact multiplanet systems with a small period ratio between neighboring planets, or near resonance. Although detecting a TTV signal certainly favors larger masses, the orbital stability of compact multiplanet systems like Kepler-11 require smaller masses and/or eccentricies given their neptune-like planetary radii. On a practical note, a compact configuration with high multiplicity reduces the risk of intermediate, non-transiting planets confusing TTV dynamical models.
 
KOI-152 (with a \textit{Kepler} magnitude 13.9, located at RA = $20^h02^m04.11^s$, Dec = $44^{\circ}22^{m}53.7^{s}$ \citealt{ste10}) is an excellent candidate for TTV modeling. It has four planetary candidates near first order Mean Motion Resonances, with clear evidence of transit timing variations (TTVs). \citet{ste12a} first noted TTVs at KOI-152 following 6 quarters of data, though only three candidates were known at the time. These three are near the 1:2:4 resonance suggesting that multi-planet resonances are reasonably common \citep{wan12}. \citet{wu13} noted the TTVs at KOI-152, and also found evidence that the inner two planetary candidates have significant eccentricities. This suggests that circular fits to the transit times would be hampered by the mass-eccentricity degeneracy (\citealt{lith12,wu13}), casting uncertainty on measured masses. 

In Section 2, we introduce our methodology for measuring transit times, and in Section 3, we examine the transit timing variations, evaluating the applicability of zero-eccentricity analytical solutions for the planetary masses. In Section 4 we describe our numerical models and fits to the transit timing variations, and present our results for the planetary masses and orbital parameters, and stellar parameters. Our analysis confirms that the candidates of KOI-152 are planets, and we refer to them as such for the remainder of the paper. In Section 5 we consider the potential effects of non-transiting perturbers on our four-planet model. In Section 6 we characterize the planets' masses, radii and by bulk densities, and in Section 7, we compare our results for KOI-152's planets with other known plamets of masses less than 30 $M_{\oplus}$ on the mass-radius diagram. 
\section{Measurement of Transit Times from \ik Photometric Time Series}
Variations in the brightness of KOI-152 were monitored with an effective duty cycle exceeding 90\% starting at barycentric Julian date (BJD) 2454964.512, 
 with all data returned to Earth at a cadence of 29.426 minutes (long cadence, LC); data were also returned at a cadence of 58.85 seconds (short cadence, SC) beginning from BJD 2455093.216. 
 Here and throughout we base our timeline for transit data from T = JD-2,454,900. Our analysis uses short cadence data where available, augmented by the long cadence dataset primarily during the epoch prior to T$ < 193$ days, for which no SC data were returned to Earth. We obtained these data from the publicly-accessible MAST archive at http://archive.stsci.edu/kepler/ . To measure the transit times from the light curve, we adopt the procedure explained in detail in Appendix 7.1 of \citet{liss13}. Here and throughout we refer to the planets as `b' (KOI-152.03), `c' (KOI-152.02) , `d' (KOI-152.01), and `e' (KOI-152.04) in order of their orbital periods. 
\section{Analytics}
We begin with the orbital periods based on a linear fit to the observed transit times, summarized in Table~\ref{tbl-intro}. We shall solve for the orbital parameters of these planets at T = 780.0 days, an epoch chosen to be near the middle of our dataset. For each candidate, the first transit time after this chosen epoch, calculated from a linear ephemeris to the set of transit times is at time $T_0$.
 \begin{table}[h!]
  \begin{center}
    \begin{tabular}{||c|c|c||}
      \hline
   \hline
      Planetary candidate  & Period (days) &  T$_{0}$ \\
\hline
 b $\;$    KOI-152.03  & 13.4845 &  784.3010 \\
 c $\;$    KOI-152.02 & 27.4023 &  806.4999 \\
 d $\;$    KOI-152.01 & 52.0909 &  821.0171 \\
 e $\;$    KOI-152.04 & 81.0631 &  802.1119 \\
      \hline
        \hline
    \end{tabular}
    \caption{A linear fit to sixteen quarters of \textit{Kepler}'s observed transit times for the planets at KOI-152, specified as orbital periods and the first calculated transit time after T = 780 days.}\label{tbl-intro}
  \end{center}
\end{table}

This configuration of planetary orbits lies close to a 1:2:4:6 resonance chain of orbital periods, and this system is known to exhibit TTVs (\citealt{ste12a,wu13,rel13,maz13}). Following the convention of \citet{lith12} and \citet{wu13}, we can measure the proximity, $\Delta$, of each adjacent pair in this chain to the nearest first order ($j:j-1$) resonance as follows: 
\begin{equation}
\Delta_1 = \frac{P'}{P}\frac{j-1}{j}-1,
\label{lith-Delta}
\end{equation}
where $P$ and $P'$ are the orbital periods of the inner and outer planets respectively.
 The expected TTV period in this case is:
\begin{equation}
P_{TTV} = \left| \frac{j}{P'}-\frac{j-1}{P}\right|^{-1}. 
\label{TTVperiod}
\end{equation}
We seek a similar measure for proximity to second order resonances ($\Delta_2$), where the expected TTV period replaces $j-1$ with $j-2$ in Equations~\ref{lith-Delta} and~\ref{TTVperiod}. Table~\ref{tbl-resonances} highlights the proximity of each pair to first or second order resonances. Each pair is close to a first order resonance (either 2:1 or 3:2). More distant pairings are close to high order (weaker) resonances, the lowest of which is the near 3:1 resonance between c and e. 

\begin{figure}[h!]
\includegraphics [height = 1.6 in]{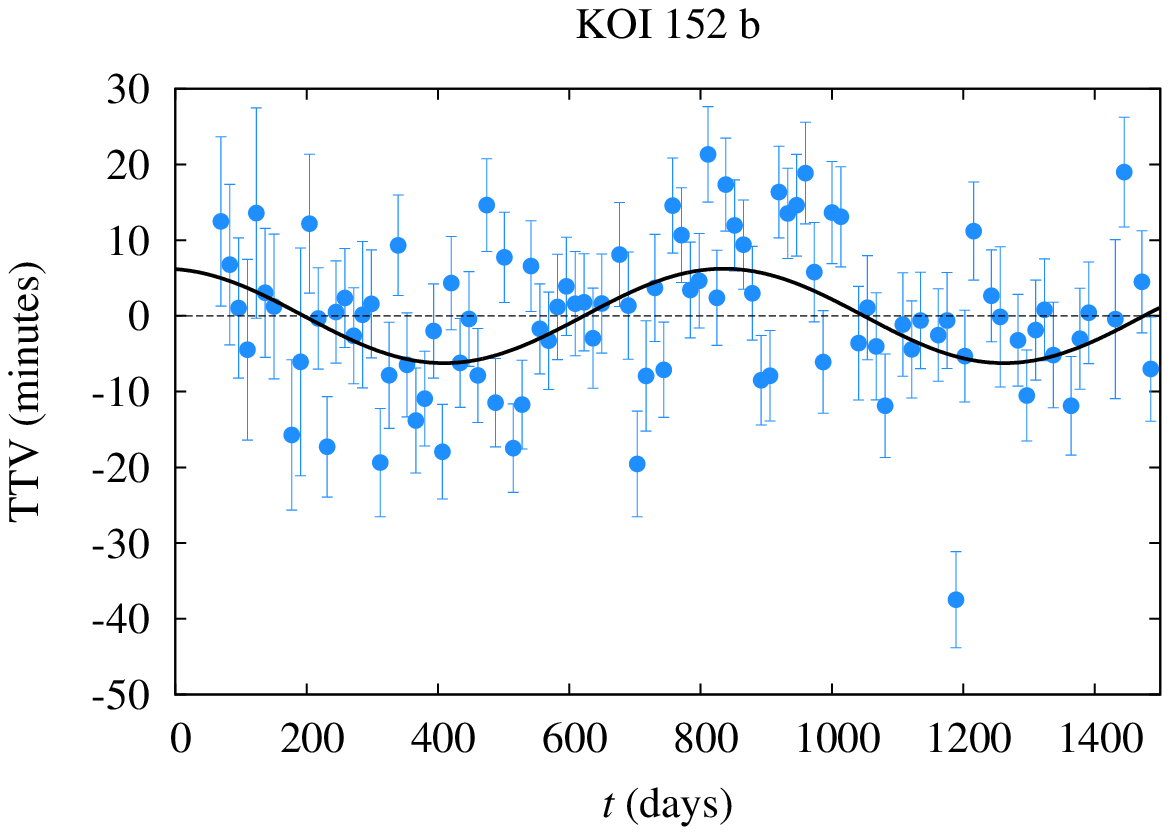}
\newline
\includegraphics [height = 1.6 in]{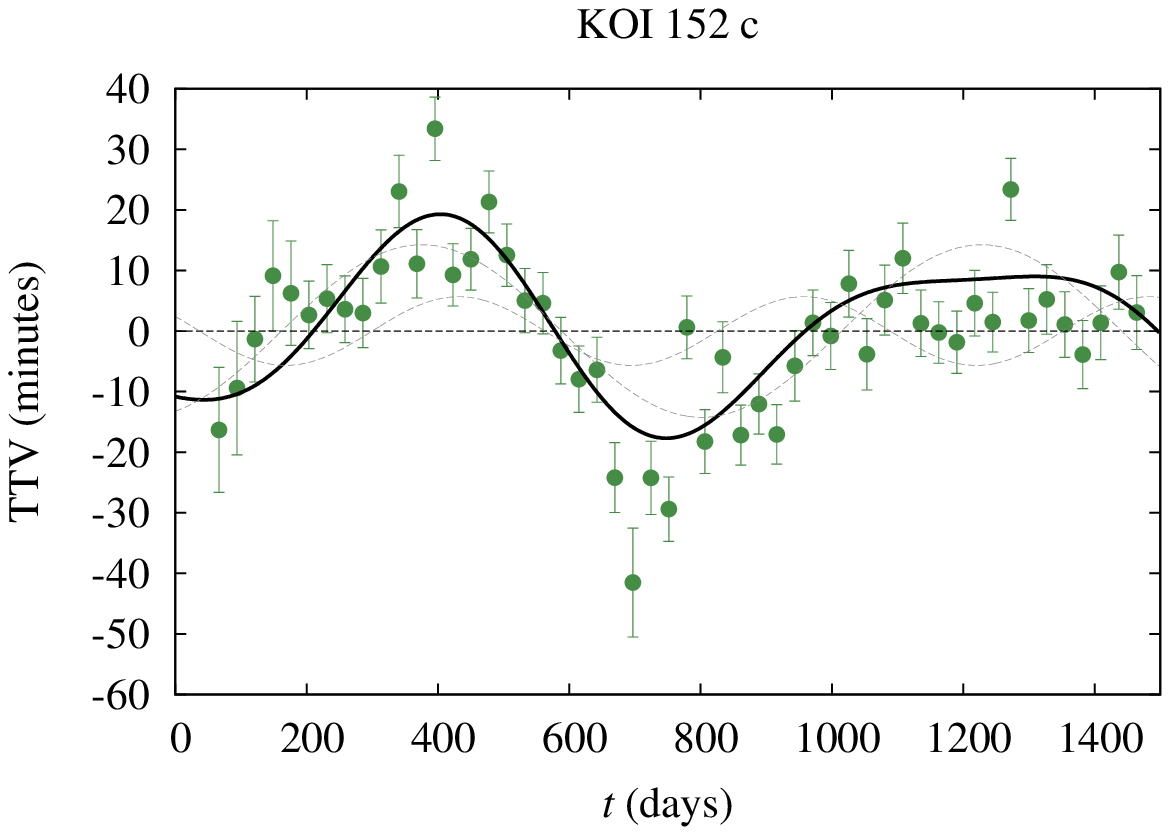}
\newline
\includegraphics [height = 1.6 in]{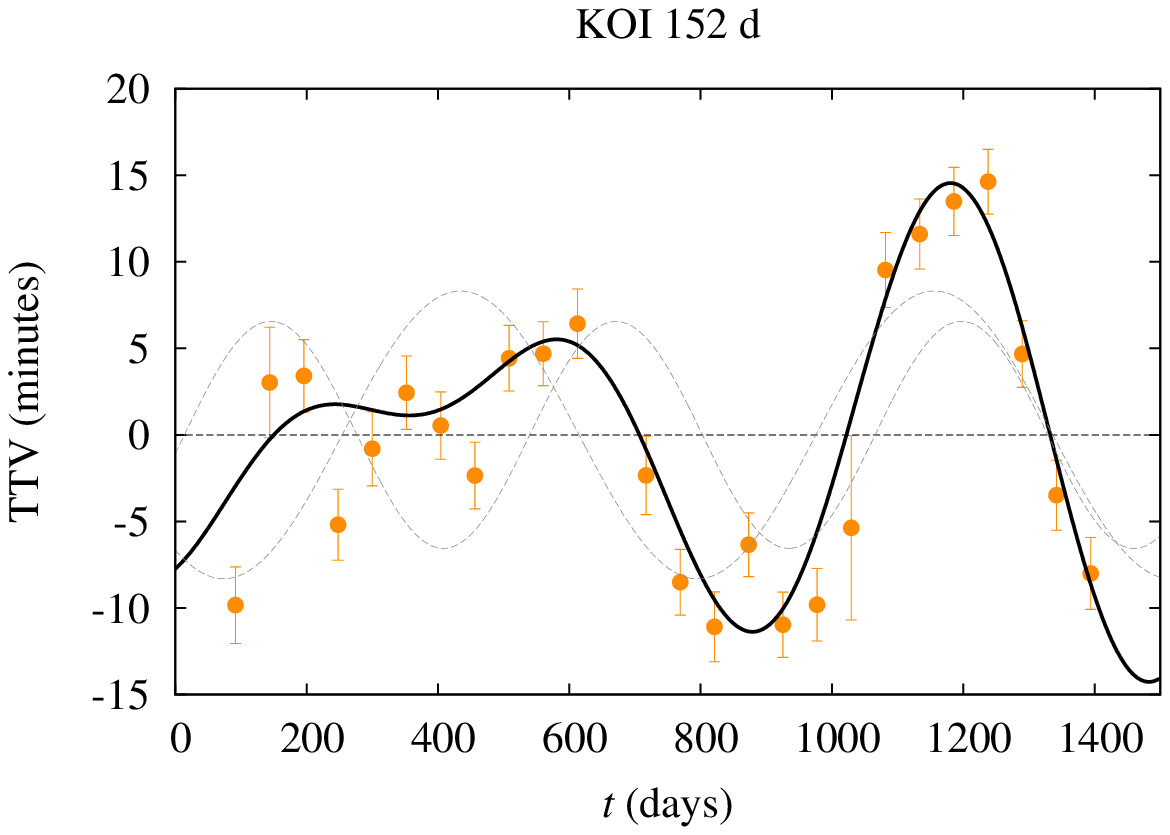}
\newline
\includegraphics [height = 1.6 in]{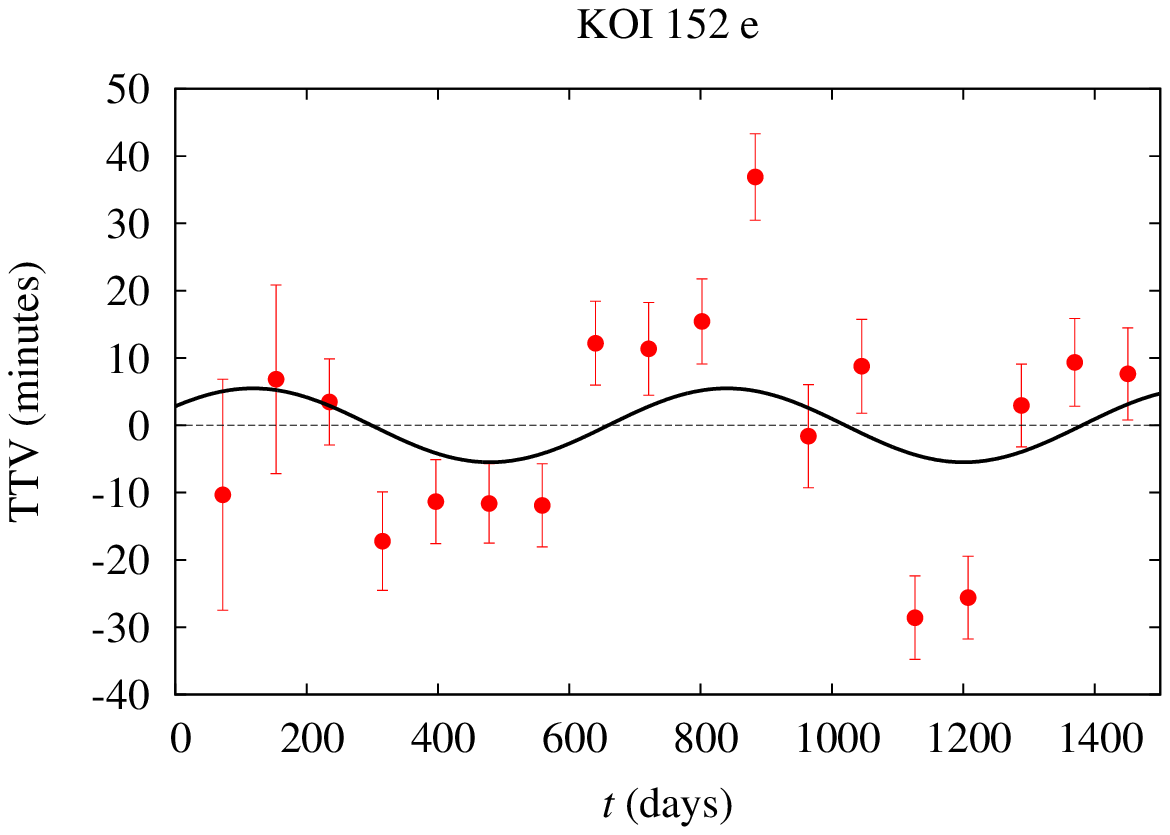}
\caption{Sixteen quarters of transit timing variations for KOI-152 (colored points), using primarily Short Cadence data, supplemented by Long Cadence data where Short Cadence was unavailable. The TTVs are the difference between the observed transit times and a calculated linear ephemeris (O-C). The solid curves are best fit sinusoids (`b' and `e'), or the sum of two sinusoids (`c' and `d': shown as dashed curves). The sunusoidal fits solve for amplitude and phase, whilst the periods remain fixed at the expected TTV period, given in Table~\ref{tbl-resonances}.}
\label{fig:1} 
\end{figure}

\begin{table}[h!]
  \begin{center}
    \begin{tabular}{||l|l|l|l|l||}
   \hline
pair & period ratio & $\Delta_{1}$ & $\Delta_{2}$ & Expected TTV period (days) \\
\hline
b,c  &  2.032 &   0.016 &  --      & 852.8     \\
c,d  &  1.901 &  -0.050 &  --      & 525.9     \\
d,e  &  1.557 &   0.037 &  --      & 721.4     \\
c,e  &  2.959 &  --     &   -0.014  & (1942.0)  \\
       \hline
        \hline
    \end{tabular}
    \caption{Orbital period ratios in the KOI 152 system, and their proximity to first order ($j$:$j-1$, third column) and second order (fourth column) resonances. The final column denotes expected TTV periodicities for each pair of potential interactions, with the periodicities near resonances that are not first order, and thus likely to produce weak perturbations in parentheses.}\label{tbl-resonances}
  \end{center} 
\end{table}
Note that for comparable TTV amplitudes, and assuming the TTVs are linear, the TTVs on `c' and `d' are likely a superposition of the two periodicities caused by their immediate inner and outer neighbors. Assuming there are no unseen perturbers, we seek to assess a model for the observed transit times to each transiting planet as as the sum of perturbations of its nearest neighbors. In Fig.~\ref{fig:1}, we show the transit timing variations for each transiting planet, and solve for sinusoidal fits to the TTVs. The TTV periods are fixed at their expected values based on the orbital periods, and the best-fit amplitudes and phases are solved by MCMC.   
The solutions to the amplitudes and phases are in Table~\ref{tbl-phases}. \citet{lith12} define the phase of the TTV for circular orbits $\phi_{ttv} = 0$, when the TTV crosses from above to below the linear fit to the orbit period. For KOI-152, we calculate time $T_{\phi}$ that TTVs transition from negative to positive for inner planets ($\phi_{ttv} = 180^{\circ}$), and where the transition is from positive to negative ($\phi'_{ttv} = 0^{\circ}$) for outer planets of the same pair. In this notation, anti-correlated TTVs from pairwise interactions on circular orbits have the same $T_{\phi}$. The uncertainties were measured by recording all extrema in TTV phase and amplitude for models within one reduced $\chi^2$ unit from the best fit solution. The uncertainties were largely symmetric and hence we quote the average of positive and negative uncertainties for simplicity. 
\begin{table}[h!]
  \begin{center}
    \begin{tabular}{||c|c|c|c|c|c||}
      \hline
   \hline
      Candidate  & TTV period (days)  & TTV ampl. (mins) &  T$_{\phi}$ (days) & T$_{\phi, circ}$  & $\Delta \phi_{ttv}$ \\
\hline
\hline
 b  & 852.8  & 6.24$\pm 1.35$ & 231.0$\pm 30.4$ & 329 & $41 \pm 13 ^{\circ}$  \\
\hline
 c  & 852.8 & 14.24 $\pm1.59$ &  264.5$\pm 14.3$ & 329  & $27 \pm 6 ^{\circ}$   \\
    & 525.9 & 5.67$\pm 1.48$  &  751.9$\pm23.0$ &  846  & $64 \pm 16^{\circ}$   \\
\hline
 d  & 525.9 & 6.56$\pm 0.96$  &  775.0$\pm 13.1$  & 846  & $49 \pm 9 ^{\circ}$   \\
    & 721.4 & 8.31$\pm 0.52$  &   467.2$\pm 14.2$ & 546  & $39 \pm 7^{\circ}$   \\
\hline
 e  & 721.4 & 5.49$\pm 1.36 $ &  422.6$\pm 31.4$ & 546  & $62 \pm 16^{\circ}$   \\
      \hline
        \hline
    \end{tabular}
    \caption{The superposition of TTVs at KOI-152, with their expected TTV periods (second column), and their best fit amplitudes (third column) and times that the TTVs transition across zero (fourth column) as depicted graphically in Fig.~\ref{fig:1}, here with measured uncertainties. The fifth column lists the times that the TTVs would transition across zero from purely circular orbits, where the longitude of conjunctions crosses the transit line of sight, and the final column measures the phase difference in degrees. All times are measured in days unless otherwise indicated.}\label{tbl-phases}
  \end{center}
\end{table}
Note that the amplitude of the TTVs for the inner two candidates are slightly less than than the measured values of \citet{wu13}, and agree within 1$\sigma$ uncertainties. For orbits with no free eccentricity, TTV phases are anti-correlated. The phases in the fourth column of Table~\ref{tbl-phases} are anti-correlated and agree at the 1.0$\sigma$ level, between `b' and `c', as well as between `c' and `d'. However, $T_{\phi}$ of `d' and `e' are separated by 1.3$\sigma$, weakly suggestive of a phase shift in their TTVs that is not equal to $180^{\circ}$.

We can perform another test on orbital phases by comparing these times $T_{\phi}$ with the date that the longitude of conjunctions is closest to the line of sight, for pairs of candidates, assuming circular orbits. These dates are in the fifth column of Table~\ref{tbl-phases}. For `b' and `c', the nearest conjunction to $T_{\phi}$ occurs at T= 329 days, for `c' and `d' at T = 846 days, and for `d' and `e' at T = 546 days. In each of these cases, there is a phase shift in the expected phase from circular orbits to the observed TTV phases. The phase shift is evidence of significant free eccentricity in the planetary orbits. 

Before we relax the assumption of circular orbits, we can calculate a quick estimate of the planetary masses. Nominal estimates of the masses of these planets can be made following the solutions of \citet{lith12,wu13}, assuming the orbits are circular. For an interacting pair of planets of mass $m$ and $m'$, orbiting a star of mass $M_{\star}$, near a ($j$:$j-1$) resonance at periods $P$ and $P'$ respectively
\begin{equation}
m =  M_{\star}\left|\frac{V'\Delta}{P'g}\right|\pi j 
\label{lith1}
\end{equation}
\begin{equation}
m' = M_{\star}\left|\frac{V \Delta}{Pf}\right|\pi j^{2/3}(j-1)^{1/3},
\label{lith2}
\end{equation}
 where $f(\Delta) =  $ and $g(\Delta)$ are numerical coefficients of the disturbing function near resonance, calculated in \citet{lith12}. Here $V$ and $V'$ are the TTV amplitudes. For planets `c' and `d', we calculate two estimates of the mass. We table these nominal masses in Table~\ref{tbl-nominalmasses}:
\begin{table}[h!]
  \begin{center}
    \begin{tabular}{||c|c||}
      \hline
   \hline
      Planetary candidate  & $m_p$ ($10^{-6}M_{\star}$) \\
\hline
\hline
 b  & 98.7$\pm$11.0 \\
\hline
 c  & 22.3$\pm$4.8 \\
    & 44.5$\pm$6.5 \\
\hline
 d  & 27.3$\pm$7.1 \\
    & 7.3$\pm$1.8   \\
\hline
 e  & 19.1$\pm$1.2  \\
      \hline
        \hline
    \end{tabular}
   \caption{Nominal mass estimates \textit{assuming circular orbits} for the planets of KOI-152, relative to the host star. For `c' and `d', perturbations on two neighboring planets permits two estimates of masses, at the amplitudes calculated in Table~\ref{tbl-phases}. Their inconsistency here may be due to significant orbital eccentricities. Our final estimates for the planetary masses are in Section 4.}\label{tbl-nominalmasses}
  \end{center}
\end{table}
Note that for both `c' and `d', the nominal masses estimated by the TTVs induced on their inner and outer neighbors are inconsistent. This implies that either the TTVs are caused by unseen perturbers, or that circular orbits are a poor fit to the data. The expectation of significant free eccentricity in the inner two planets of KOI-152 was noted by \citet{wu13}. Due to the degeneracy between mass and eccentricity in TTVs, the expectation of eccentricity casts doubt upon planetary mass estimates made under an assumption of circular orbits.

 TTVs with significant non-sinusoidal components contain information that is difficult to probe analytically. Hence we perform numerical fits to the transit times to solve for masses and the osculating orbital parameters of the planets at the epoch T$= 780$ days.
\section{Dynamical models of KOI-152 with four planets}
\begin{figure}[placement here]
\includegraphics [height = 1.9 in]{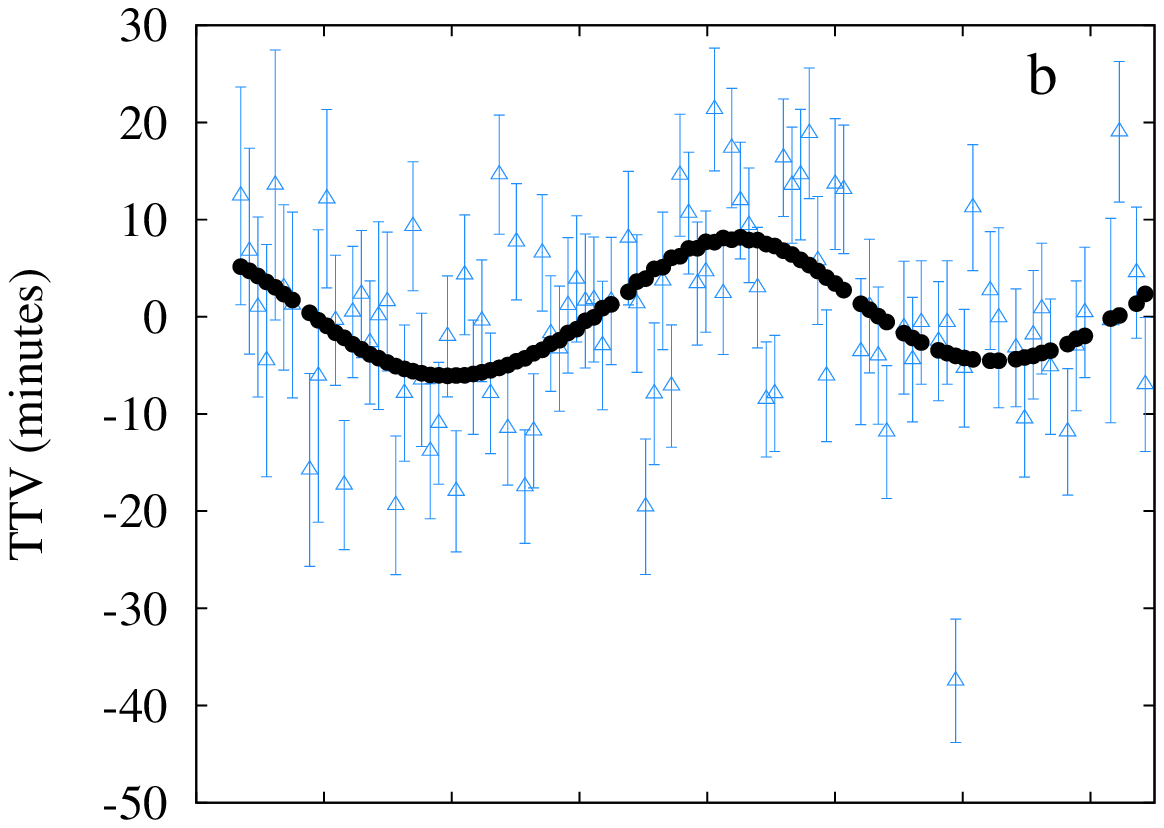}
\includegraphics [height = 1.9 in]{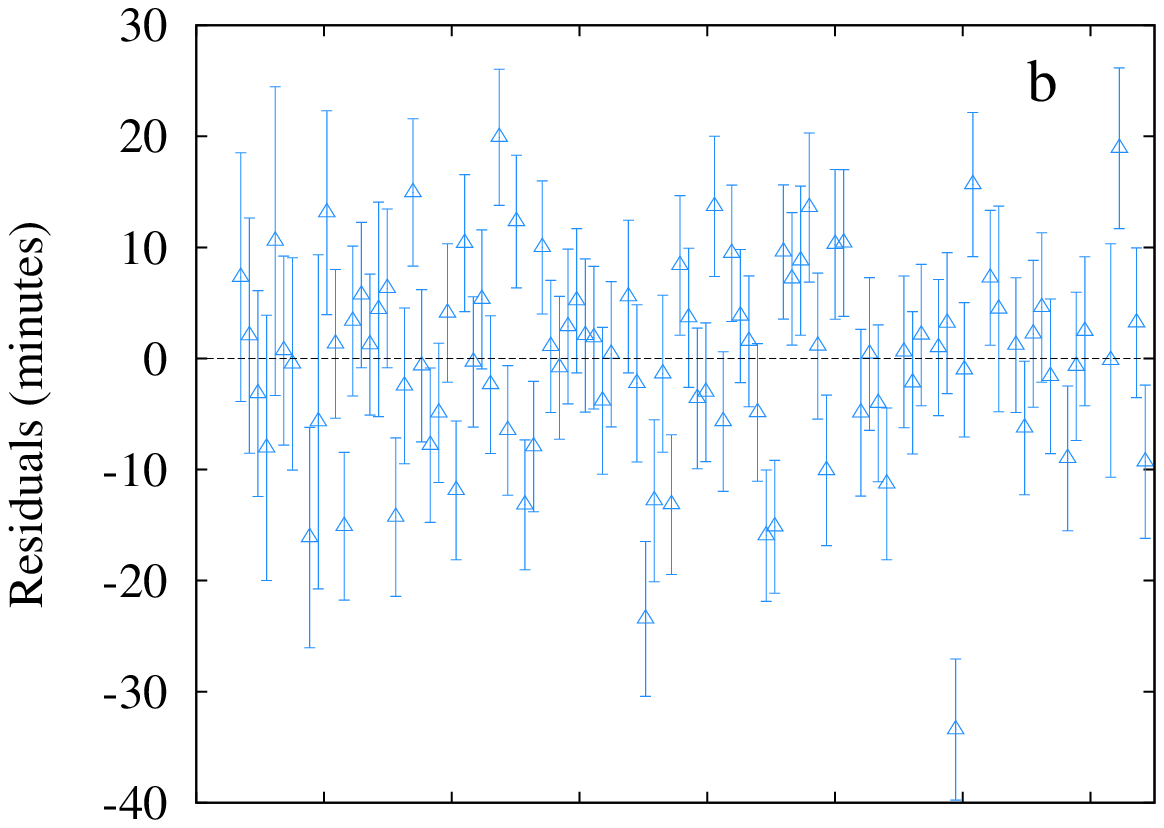}
\newline
\includegraphics [height = 1.9 in]{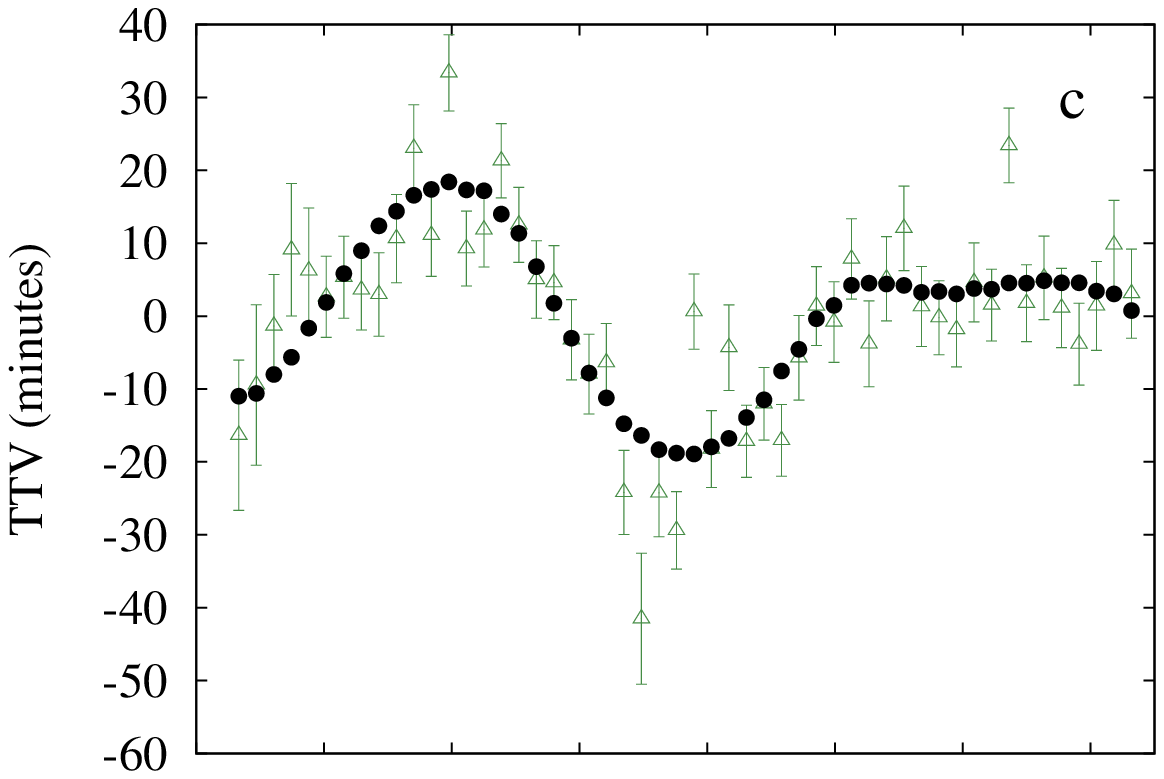}
\includegraphics [height = 1.9 in]{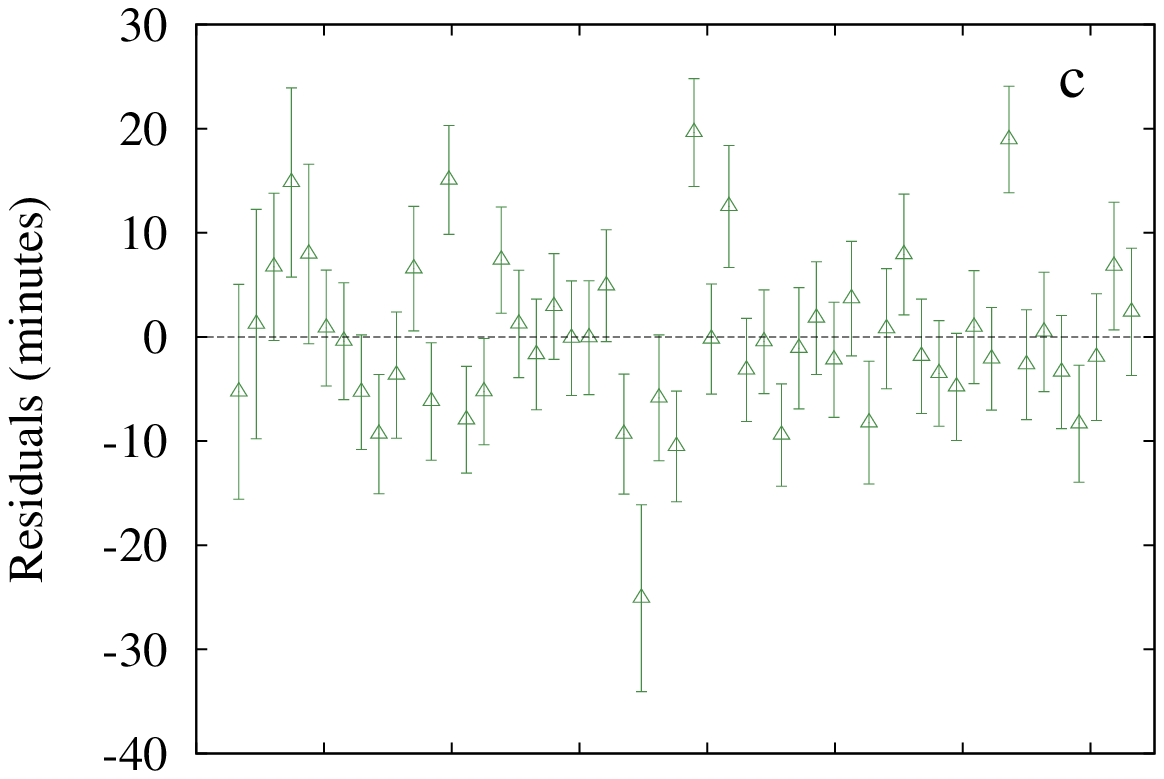}
\newline
\includegraphics  [height = 1.9 in]{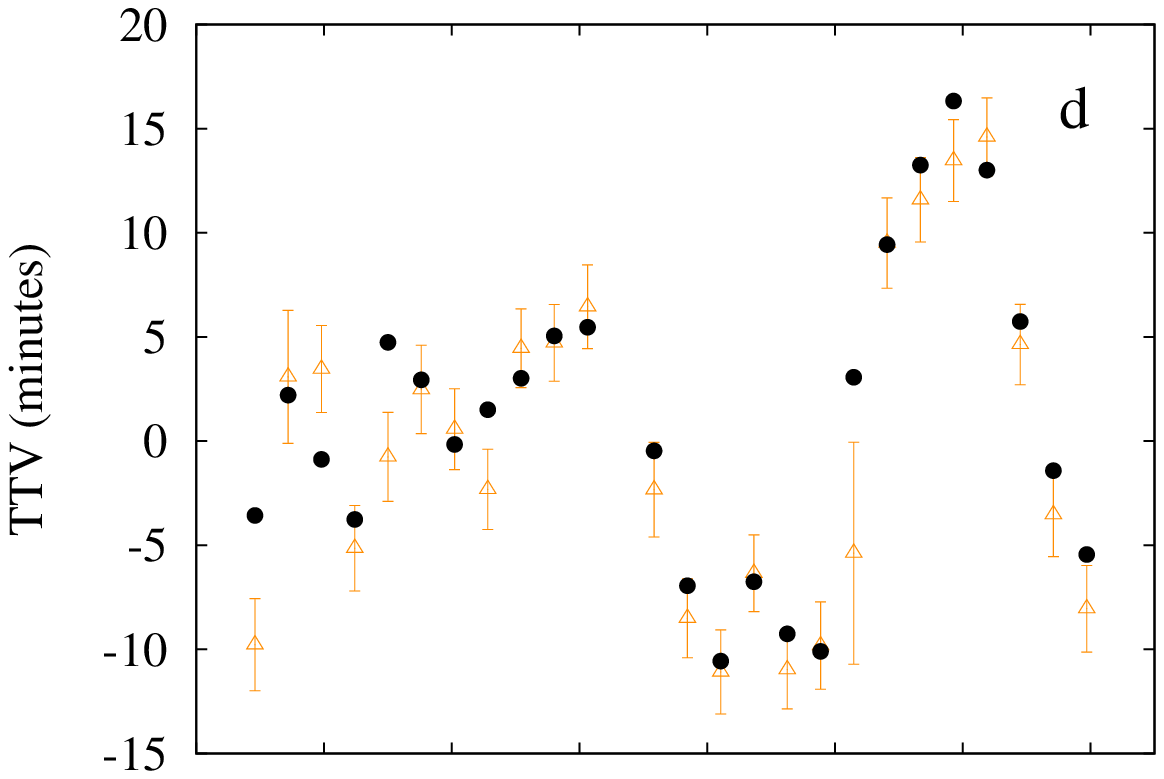}
\includegraphics [height = 1.9 in]{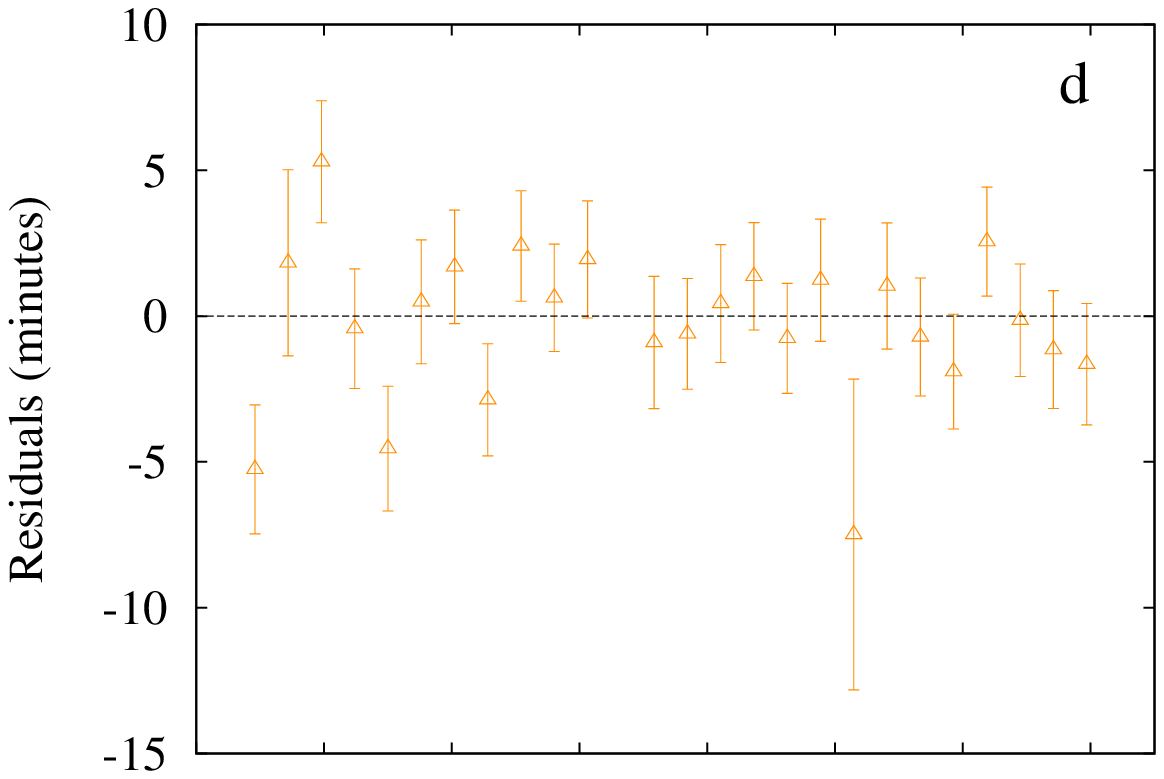}
\newline
\includegraphics  [height = 1.9 in]{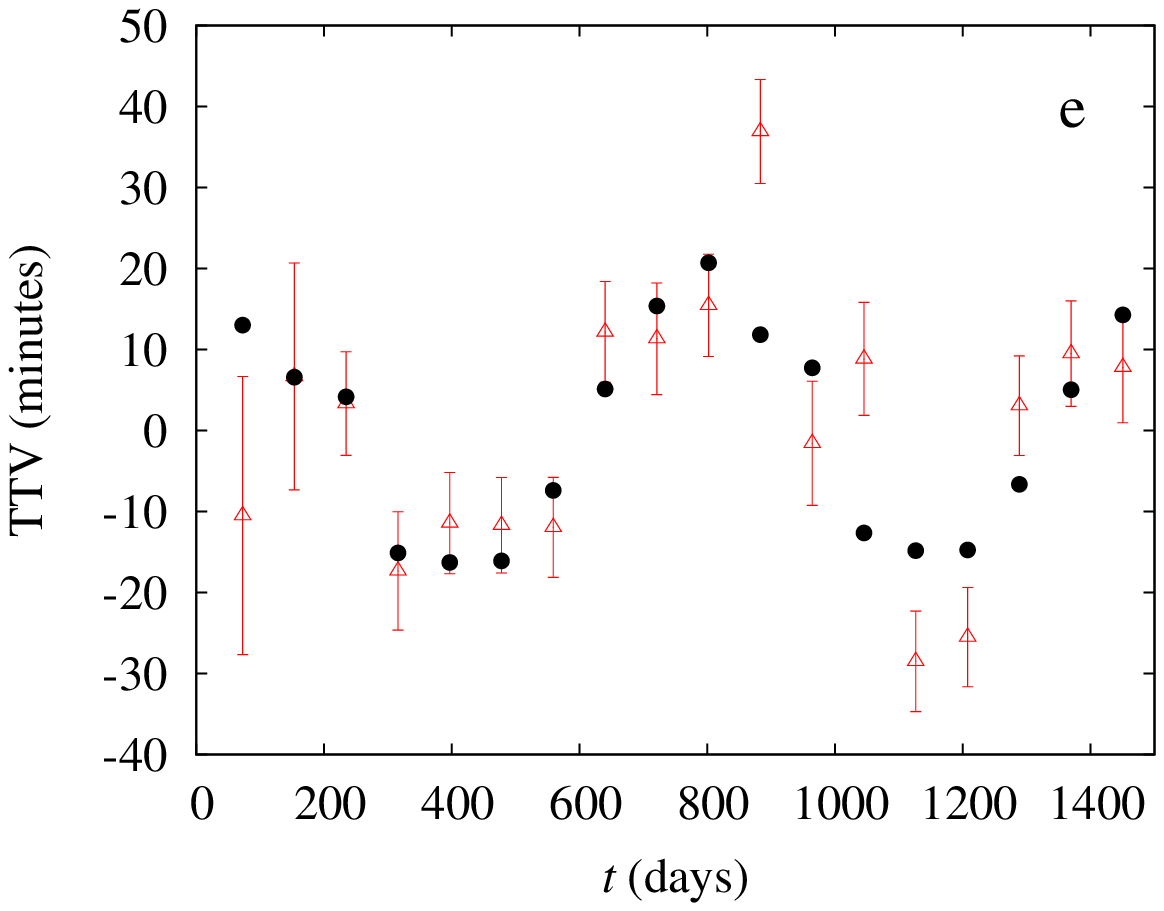}
\includegraphics [height = 1.9 in]{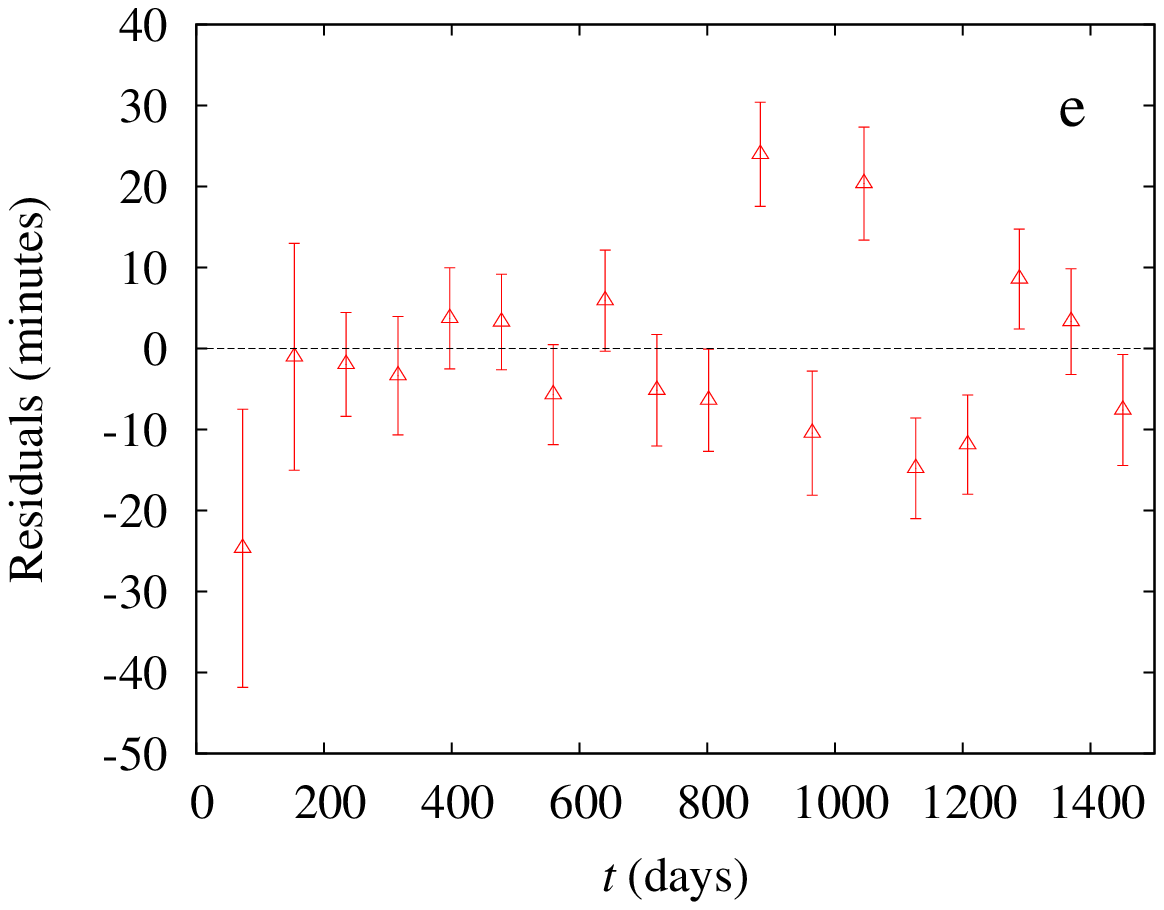}
\caption{Observed and simulated transit timing variations for the planetary candidates orbiting KOI-152, using the combined dataset of short and long cadence transit times. The panels on the left side compare $O-C$ (colored data) and model times, $S-C$  (black points). The right hand side plots the residuals with the dynamical model subtracted from the observed transit timing variations. Note the vertical scales for the residuals (right panels) differ from those of the TTVs.}
\label{fig:2} 
\end{figure}

Figure~\ref{fig:2} shows the transit timing variations for the four known candidates and a dynamical fit to this dataset. Our free parameters were orbital period, the time of the first transit $T_{0}$ after epoch ($T= 780$ days), the eccentricity vectors $e\cos\omega$ and $e\sin \omega$, (where $\omega$ is measured from the sky-plane and reaches 90$^{\circ}$ if the pericenter coincides with the transit line of sight), and planetary mass. The dynamical models measure mass as a fraction of the stellar mass. However, following \citet{liss13}, an accurate constraint on $\rho_{\star}$ and hence the stellar mass is one of the benefits of dynamical models. To integrate planetary motions, we adopt the 8th order Runge-Kutta Prince-Dormand method, which has 9th order errors. In all of our models, the orbital period and phase of each of the planets are free parameters. The phase is specified by the midpoint of the first transit subsequent to our chosen epoch. We keep all planetary masses as free parameters. We have assumed co-planarity, i.e., negligible mutual inclinations between planetary orbits, in all of our dynamical models. We make no attempt to model transit durations or impact parameters in our dynamical simulations. 

Our integrations produce an ephemeris of simulated transit times, $S$, and we compare these simulated times to the observed transit times ($O$). We employ the Levenberg-Marquardt algorithm to search for a local minimum in $\chi^2$. The algorithm evaluates the local slope and curvature of the $\chi^2$ surface. Once it obtains a minimum, the curvature of the surface is used to populate the covariance matrix and evaluate uncertainties. Other parameters are allowed to float when determining the limits on an individual parameter's error bars. Assuming that the $\chi^2$ surface is parabolic in the vicinity of its local minimum, its contours are concentric ellipses centered at the best-fit value. The orientations of these ellipses depend on correlations between parameters. The errors that we adopt account for the increase in uncertainty in some dimensions due to such correlations. 

Our best fit model for the nominal 190 transit times (shown as TTVs in Fig.~\ref{fig:1}) of the four known planetary candidates leaves 6 data points that are outliers beyond 3$\sigma$ (where $O-S/\sigma$ > 3)  (of which there should be $\approx$ 0.4, i.e., likely zero or one, if the errors were Gaussian), and 20 points that are between 2 and 3$\sigma$ (of which there should be $\approx$ 8 if the errors were Gaussian). Clearly the uncertainty in the transit times is either under-estimated or the four-planet model is wrong. The outliers may be due to errors in some of the the measured transit times that are not incorporated into timing uncertainty estimates; sources of such errors could be stellar activity, instrumental effects, etc. 

To assess our dynamical model with $\chi^2$ minimization requires a method for dealing with these outliers. \citet{liss13}, in their TTV analysis of Kepler-11, compared three independent methods of measuring transit times from light curves to filter out outlying or anomalous transit times. They discarded points that did not have overlapping error bars for transit times with at least one of the other two transit time measurements. Here, for KOI-152,  we seek a method of self-filtering a single measured set of transit times, noting that the distribution of measured times has far too many 3$\sigma$ outliers as well as too many 2$\sigma$ outliers. Hence, we compare best fit dynamical models against the combined `raw' set of short and long cadence transit times with best fit models where outliers beyond 3$\sigma$ or 2$\sigma$ respectively are removed, and a dataset of short cadence only transit times with outliers beyond 3$\sigma$ removed. We thus conduct our dynamical fits against four sets of transit times.

For each set of transit times, we find a best fit solution and then evaluate which outliers, if any, are beyond our 2$\sigma$ or 3$\sigma$ threshold. These points are removed and another best fit solution is found with dynamical models. Iterations continued until there were no more outliers at the best fit solution.

To account for multiple local minima in the $\chi^2$ surface, we used multiple choices of initial conditions, recording all solutions with $\chi^2$ values within one reduced $\chi^2$ unit of the best fit model. The outputs of each local minimum were used as initial conditions for all other sets of transit times in the search for alternative local minima in the $\chi^2$ surface. The search over different initial conditions continued until at least 25 local minima within one reduced $\chi^2$ unit of the best known fit for each of the four sets of transit times were found.

\begin{table}[h!]
  \begin{center}
    \begin{tabular}{||c|cccc||}
      \hline
Planet  &  $\chi^2_{raw}$ & $\chi^2_{3\sigma}$  &   $\chi^2_{2\sigma}$  &    $\chi^2_{3\sigma SC}$ \\
      \hline
              b & 173.27 & 107.51   & 63.79     & 102.12  \\
              c & 88.95  & 56.25   &  32.49     & 51.96 \\
              d & 30.12  & 30.61   &  21.73     & 22.73  \\
              e & 43.52  & 14.74   &  9.79     &  13.44  \\
       \hline
        total & 335.87  & 209.11  & 127.80    & 190.24  \\
$\#$ fitted TTs  &  190    &   181   & 167       & 163     \\
 $\chi^2/$(d.o.f.) & 1.98  & 1.30  & 0.87   & 1.33      \\
       \hline
    \end{tabular}
    \caption{$\chi^2$ contributions from each planet for a suite of best fit models against four sets of transit times: the raw measured transit times of short and long candence data combined (raw: second column), a combined set with 3$\sigma$ outliers ($O-S/\sigma > 3$) removed (3$\sigma$: third column), a combined set with 2$\sigma$ outliers removed (2$\sigma$: fourth column), and in the last column, short cadence only data with 3$\sigma$ outliers removed (3$\sigma$SC). The degrees of freedom (d.o.f.) in each case are the number of data points in each set of transit times (the penultimate row) subtract the number of free parameters in each fit (twenty free parameters).}
\label{tbl-fits}
  \end{center}
\end{table}
Table~\ref{tbl-fits} summarizes the goodness of fit of the best fit model for each set of transit times. Including all measured transit times leads to $\chi^2/\rm{(d.o.f.)} >1 $, whereas, not surprisingly, using the highest threshold of acceptance for the transit times ($O-S/\sigma <2$), causes $\chi^2/\rm{(d.o.f.)} < 1$. We note here that KOI-152 d contributes the least to the $\chi^2$ when all measured transit times are included, and appears to be the least sensitive to the removal of outlying transit times. 
\begin{figure}[h!]
\includegraphics [height = 2.3 in]{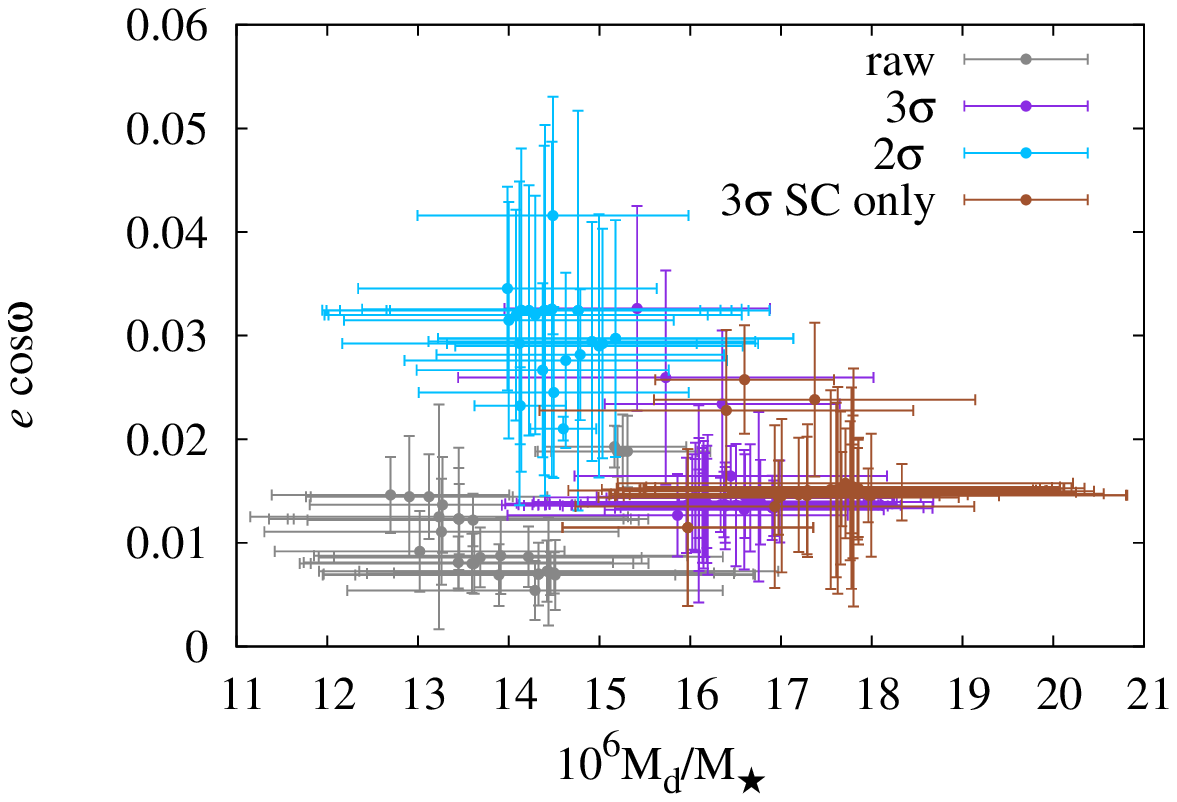}
\includegraphics [height = 2.3 in]{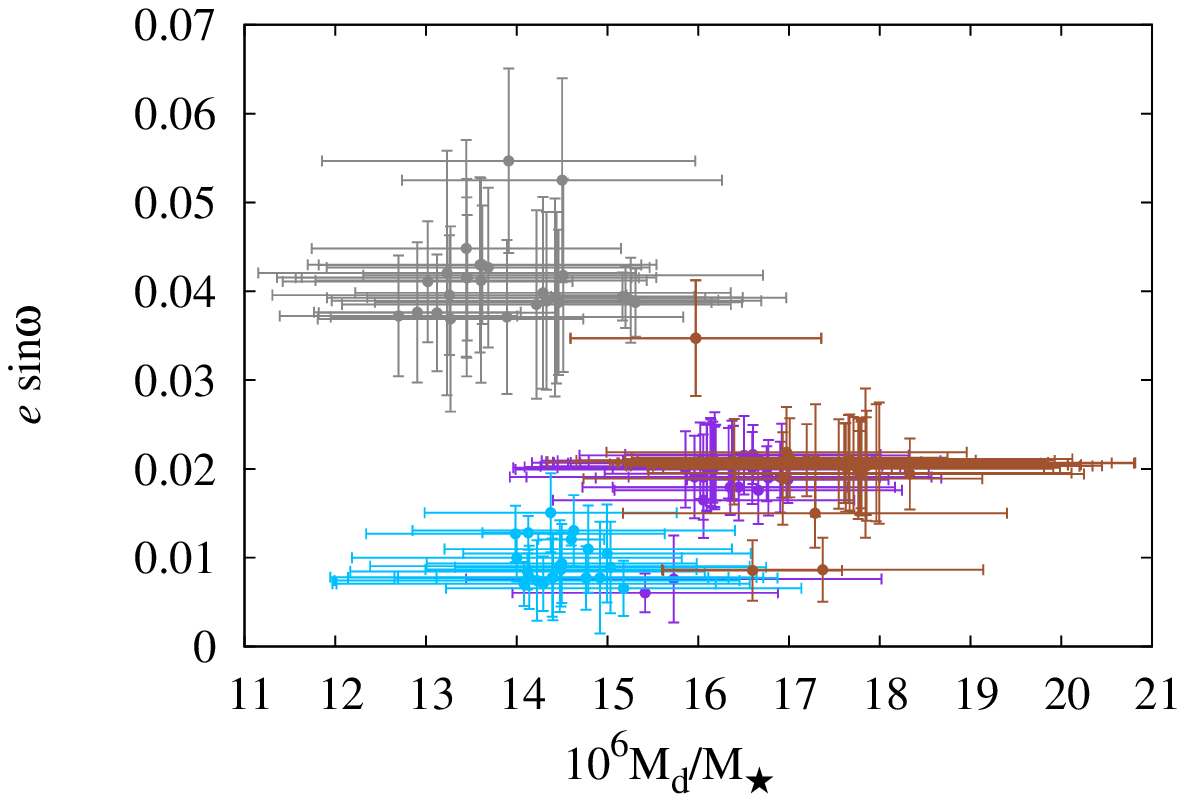}
\caption{The results of fitting best fit mass and the components of the eccentricity vector for KOI-152 d using four set of transit times. Each point represents a local minimum in $\chi^2$ with uncertainties estimated locally from the covarience matrix. For each set of transit times, local minima were included if, compared to the lowest known $\chi^2$ for that dataset, $\frac{\Delta\chi^2}{\chi^2/(d.o.f.)} < 1$, and their error bars were reduced such that with uncertainties, $\frac{\Delta\chi^2}{\chi^2/(d.o.f.)} = 1$.}
\label{fig:m-eccs-d} 
\end{figure}

Figure~\ref{fig:m-eccs-d} shows the range of masses and eccentricity vectors that are with one reduced $\chi^2$ unit of the best known minimum for each dataset of transit times. All model fits within one reduced $\chi^2$ unit of the best known fit for each dataset of transit times were included. The nominal uncertainties for model fits $\sigma_{nom}$, are reduced for all minima apart from the one with the lowest $\chi^2$, with uncertainties reduced to $\sigma_r = \sigma_{nom} \sqrt{1-\frac{\Delta \chi^2}{\chi^2/(d.o.f.)} }$. This assumes that the local minimum in $\chi^2$ is on a parabolic surface, and therefore extends error bars to reach where $\frac{\Delta \chi^2}{{\chi^2/(d.o.f.)}} \approx 1$. The result in Figure~\ref{fig:m-eccs-d} shows that a wide range of eccentricities can satisfy the data for KOI-152 d, and that individual best fit solutions are indeed moderately sensitive at the $\sim 1\sigma$ level to the manner in which outlying transit time measurements are handled. 

Nevertheless, with our exploration of multiple local minima in the $\chi^2$ surface, our uncertainties are augmented to ensure that our best fit models over all four sets of transit times. We note that within each dataset, the uncertainties with differing datasets are  all consistent at the 1$\sigma$ level.
\begin{figure}[h!] 
\includegraphics [height = 2.3 in]{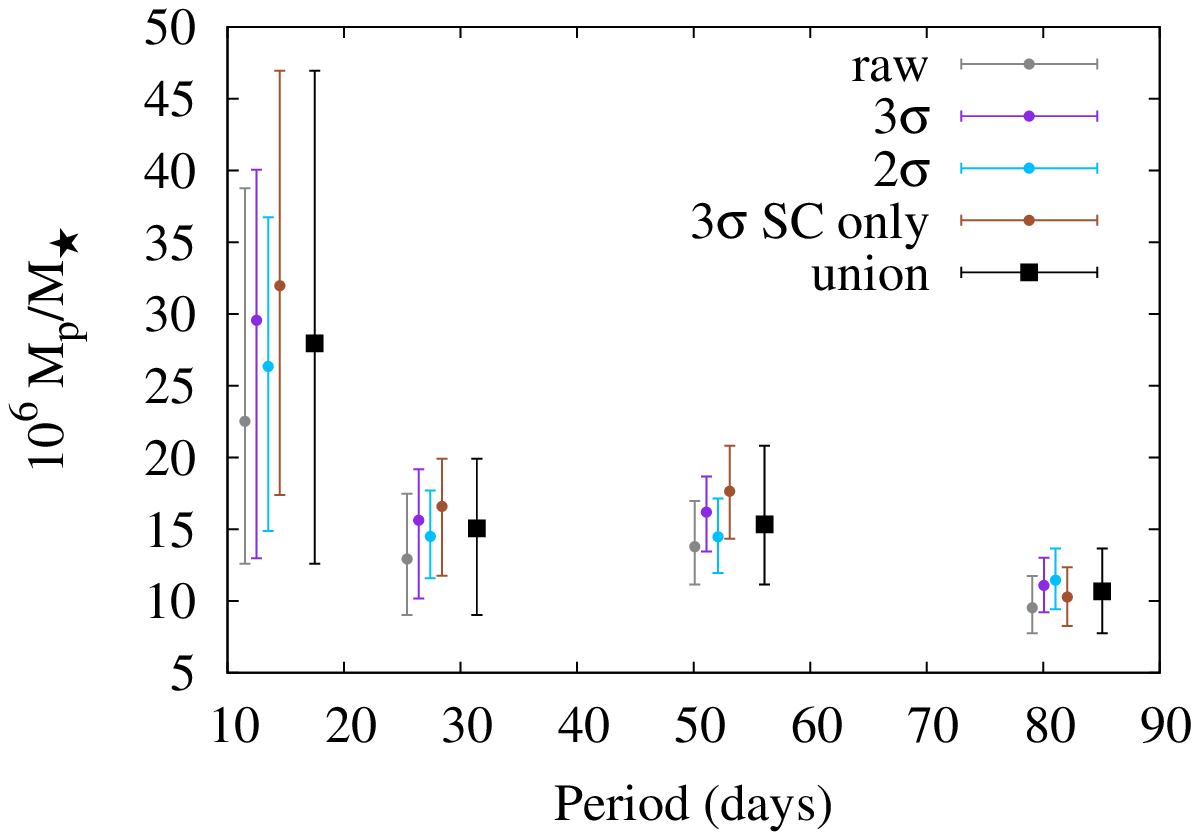}
\includegraphics [height = 2.3 in]{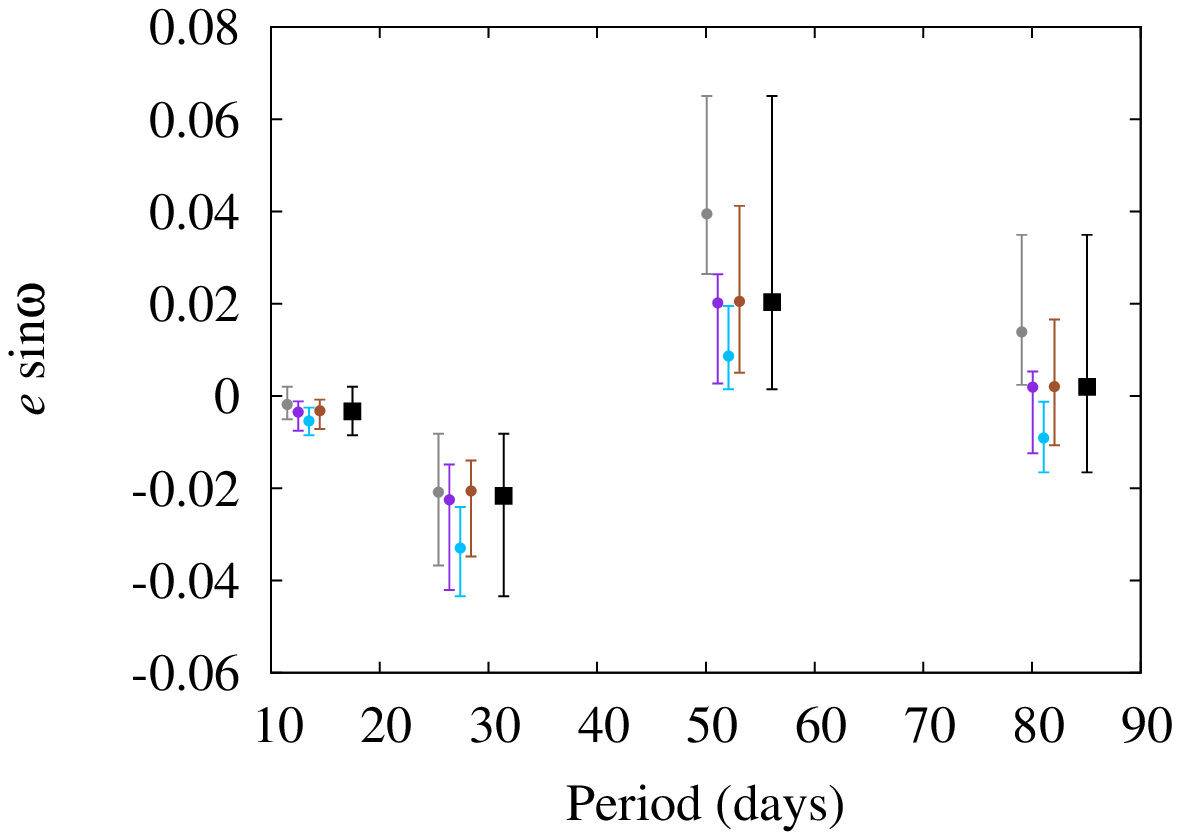}
\newline
\includegraphics [height = 2.3 in]{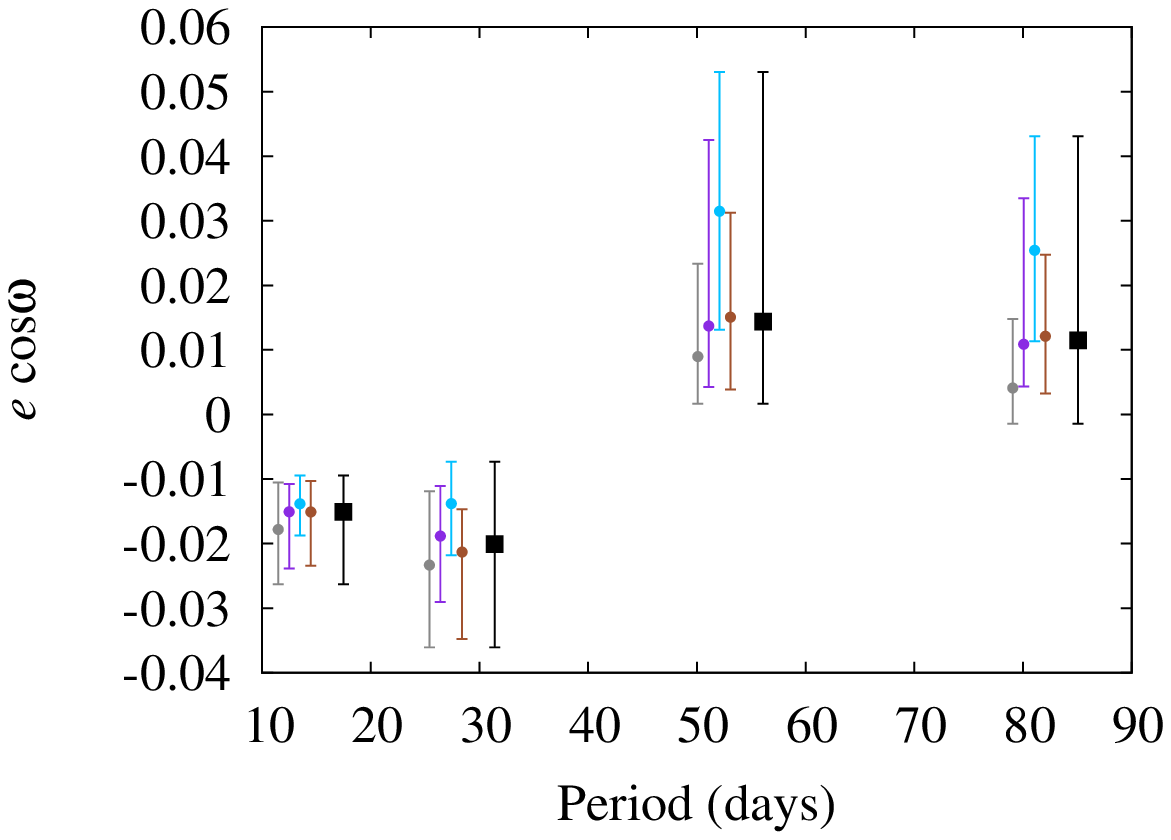}
\caption{Best fit solutions for planetary masses and the eccentricity vector components $e\sin\omega$ and $e\cos\omega$ using four different methods of accounting for outliers amongst the measured transit times. The orbital periods are displayed with a stagger to allow comparison between the solutions. Here, the dynamical fit against raw transit times is marked by the grey points, the combined dataset of short and long cadence transit times with 3$\sigma$ outliers removed is in blue, the same set with 2$\sigma$ outliers removed is in red, and the short cadence data only with 3$\sigma$ outliers removed is in green. The black squares mark the median of the four solutions, with error bars showing the union of all uncertainties.}
\label{fig:union} 
\end{figure}
In Fig.~\ref{fig:union}, we show the 1$\sigma$ uncertainties for the masses and the components of the eccentricity vectors for all four planets. For our total uncertainty for each parameter, we adopt the union of all uncertainties for each parameter as shown in Fig.~\ref{fig:union}, and the median of the four best fit solutions for each parameter as our nominal solution. The results for all the fitted parameters are in Table~\ref{tbl-solutions}.
\begin{table}[h!]
  \begin{center}
    \begin{tabular}{|cccccc|}
      \hline
      \hline
Planet     & Period (days) & T$_0$ (JD-2,545,900) & $e\cos\omega$ & $e\sin\omega$ & $10^{6}\frac{M_p}{M_{\star}}$ \\
 b &  \textbf{ 13.4845 } $^{+ 0.0002 }_{- 0.0002 }$ & \textbf{ 784.307 } $^{+ 0.002 }_{- 0.002}$ & \textbf{ -0.015 }$ ^{+ 0.006 }_{- 0.011}$ & \textbf{ -0.003 }$^{+ 0.005  }_{- 0.005}$ & \textbf{28.0 }$^{+ 19.0 }_{- 15.4}$  \\
 c &  \textbf{ 27.4029  } $^{+ 0.0008 }_{- 0.0006 }$ & \textbf{ 806.475 } $^{+ 0.004 }_{- 0.004}$ & \textbf{ -0.020 }$ ^{+ 0.013 }_{- 0.016 }$ & \textbf{ -0.022 }$^{+ 0.014  }_{- 0.022}$ & \textbf{15.1 }$^{+ 4.9 }_{- 6.0}$  \\
 d &  \textbf{  52.0902  }$ ^{+ 0.0009 }_{- 0.0010 }$ & \textbf{ 821.011 } $^{+ 0.002 }_{- 0.001 }$ & \textbf{ 0.014 }$ ^{+ 0.039 }_{- 0.013}$ & \textbf{ 0.020 }$^{+ 0.045  }_{- 0.019}$ & \textbf{15.3}$^{+ 5.5  }_{- 4.2}$  \\
 e &  \textbf{  81.0659  }$ ^{+ 0.0013 }_{- 0.0011 }$ & \textbf{ 802.126 }$ ^{+ 0.003 }_{- 0.005 }$ & \textbf{ 0.012 }$ ^{+ 0.032}_{- 0.013}$ & \textbf{ 0.002 }$^{+ 0.033 }_{- 0.019}$ & \textbf{10.7 }$^{+ 3.0 }_{- 2.9}$  \\
     \hline
       \hline
    \end{tabular}
    \caption{Our solutions for KOI-152 following dynamical fits incorporating uncertainties over four choices of observed transit times with different outliers excluded. The parameters we measure include the orbital periods (second column), time of first transit after epoch (third column), $e\cos\omega$ (fourth column), $e\sin\omega$ (fifth column), and planetary mass relative to the mass of the star (sixth column).}\label{tbl-solutions}
  \end{center}
\end{table}

To test these solutions for long-term orbital stability, we simulated the trajectories of the best fit solutions using a symplectic integrator \citep{rh02} for 400 Myr and found the system to be stable. However, increasing the masses and eccentricities to their 1$\sigma$ maxima, the system lasted just 70,000 years before planets were expelled, an indicator that, dynamically, KOI-152's planets are packed close to the stability limit.

\begin{figure}[h!]
\includegraphics [height = 1.6 in]{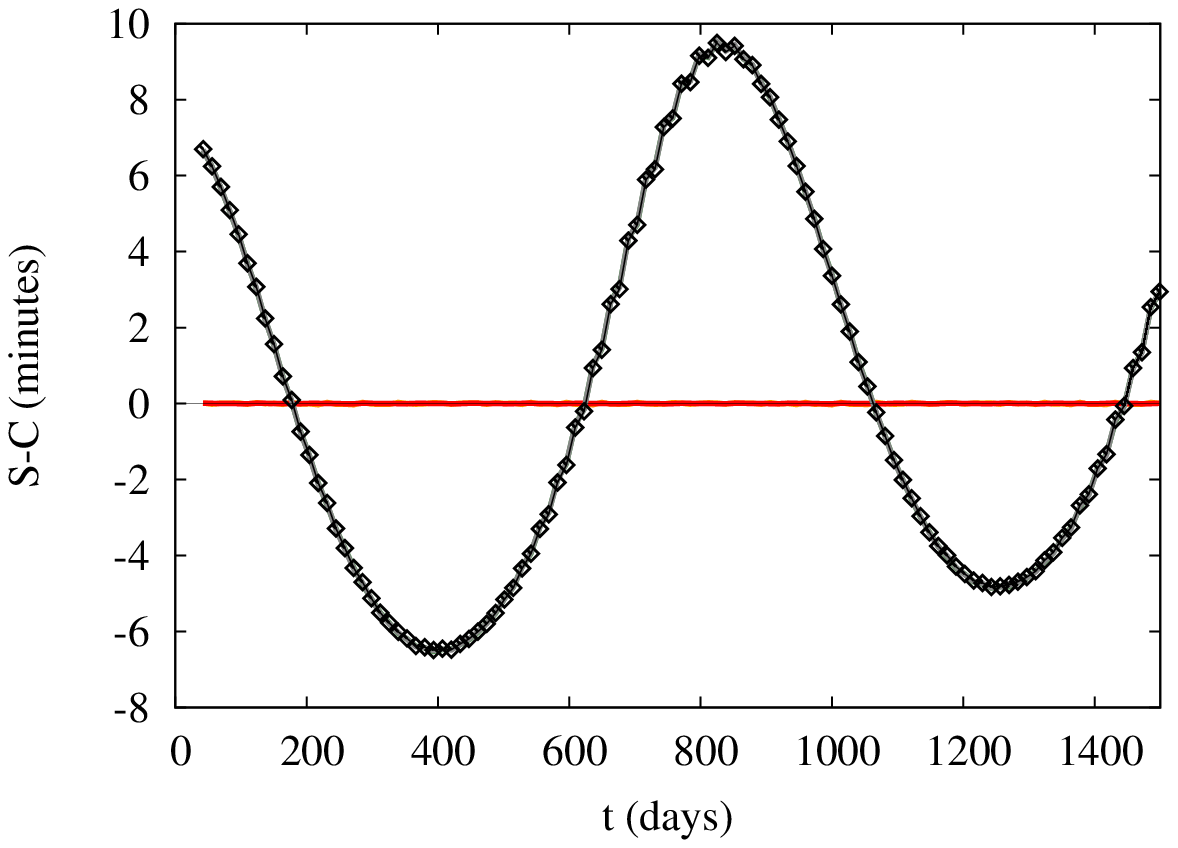}
\newline
\includegraphics [height = 1.6 in]{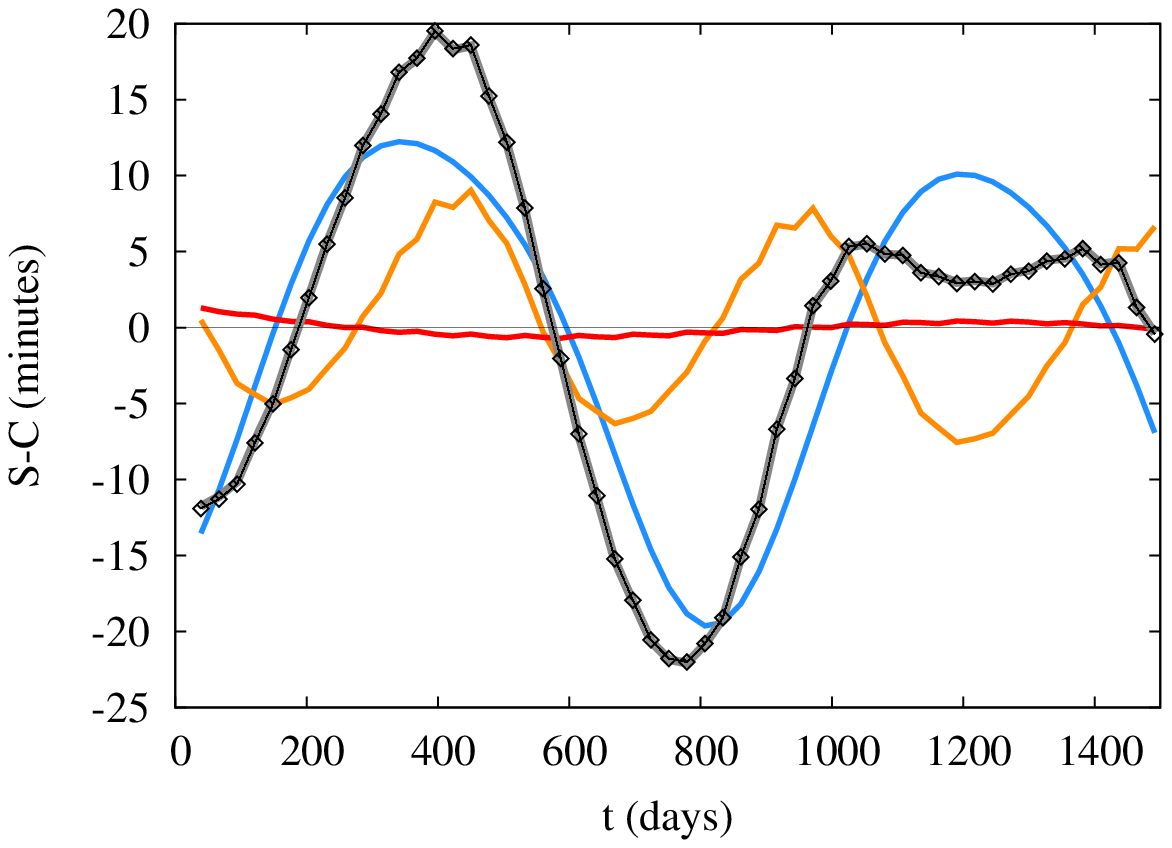}
\newline
\includegraphics [height = 1.6 in]{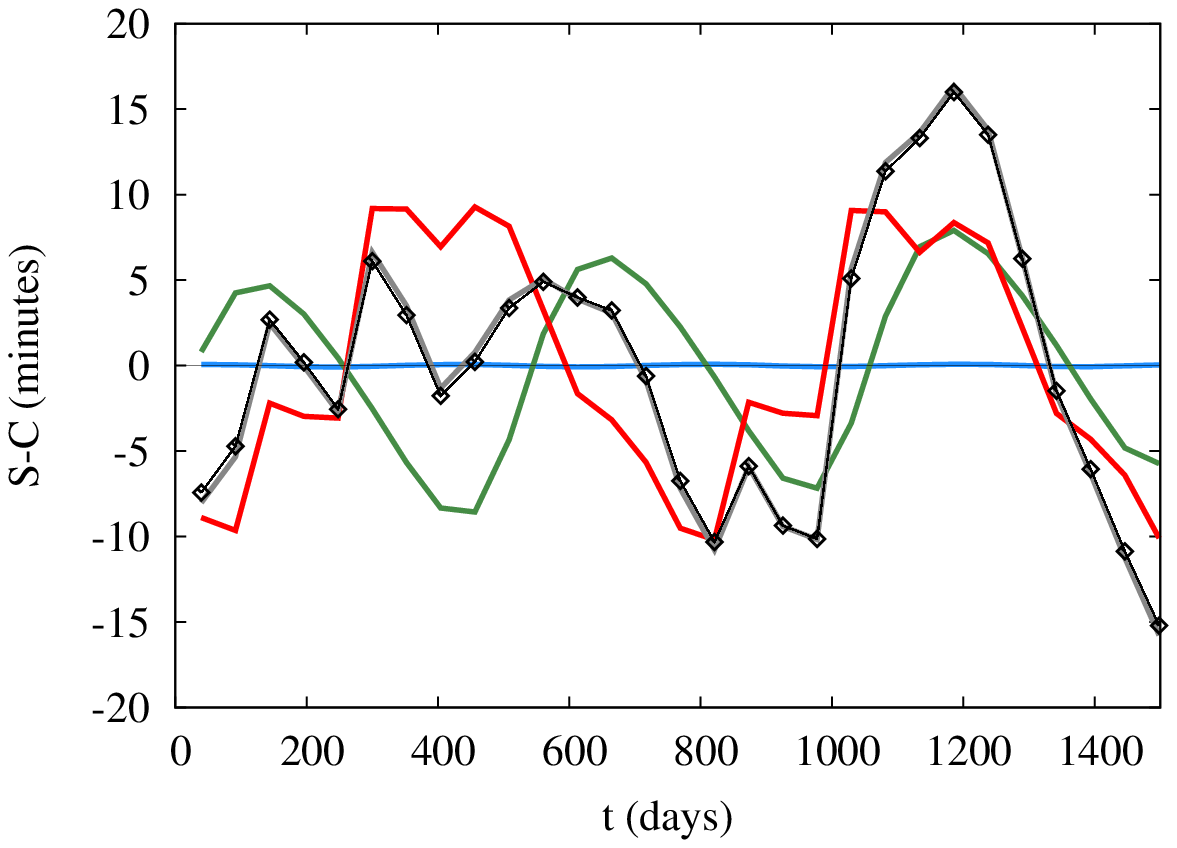}
\newline
\includegraphics [height = 1.6 in]{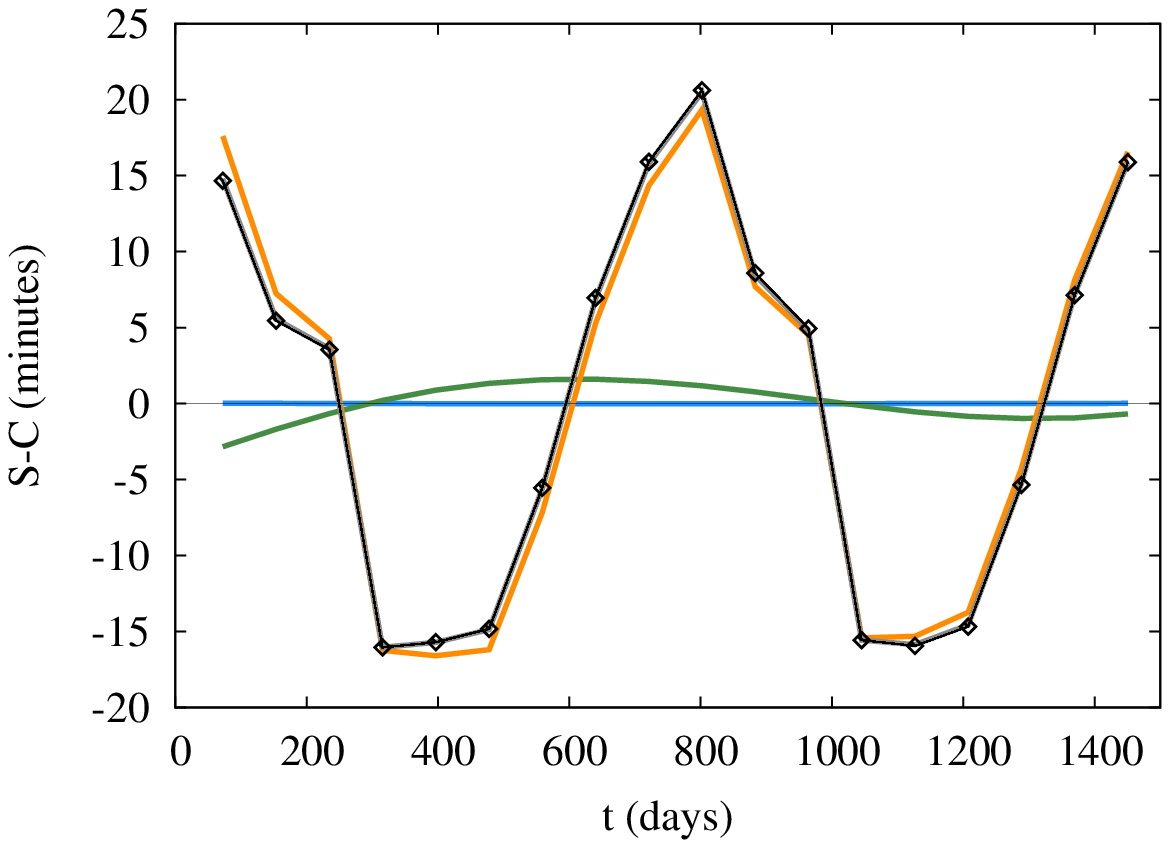}
\caption{Pairwise contributions to TTVs from each planet (simulated transit times minus a calculated linear ephemeris to the transit times) in the best fit solution to the dataset of transit times with 3$\sigma$ outliers removed for `b' (top left), `c' (top right), `d' (bottom left), and `e' (bottom right). In each panel, contributions from `b' are in blue, `c' in green, `d' in orange, and `e' in red. The contributions are summed and displayed in grey, although these curves are largely obscured by the black curves in most of the panels, which denote the best fit solution, in close agreement with the sum of linear pairwise TTVs.}
\label{fig:pairwise} 
\end{figure}

The perturbations of each planet orbiting KOI-152 can be deconvolved as a linear sum of pairwise TTVs to a high degree of accuracy. Figure~\ref{fig:pairwise} highlights the contribution of each planetary candidate to the TTVs of the other planets in the system, including the non-sinusoidal components to the TTV signal that are captured by dynamical fits. The remarkable fit of the pairwise perturbations in adding up to match the net TTVs is indicative of the linear nature of KOI-152's TTVs over the \textit{Kepler} observational baseline. The orbits of KOI-152's planets are close enough to resonance for the coherence period, the length of time over which perturbations act constructively, to be much longer than synodic period. However, the orbits are not so close to resonance that coherent perturbations reach high amplitude, and the TTVs are always a tiny fraction of an orbital period. Hence the TTVs remain in the linear regime. We also note that the near-second-order resonance TTVs on KOI-152 c induced by `e' have an amplitude of just one minute and the TTVs of KOI-152 e induced by `c', have an amplitude of just 2 minutes, even though for these orbital periods near 3:1, $|\Delta_2| = 0.014$ is lower than each of the first order nearness-to-resonance values ($\Delta_1$). Note that the expected period of this component of the TTVs is much longer than the baseline of observations and the cycle is incomplete.

To characterize the host star, we used the light curve alongside spectral classification. For fitting the light-curve, we adopt the analytic model of \citet{man02} for a planet transiting a stellar surface described by a quadratic limb-darkening law, and we adopt the limb-darkening parameters of \citet{cla11}. We modelled the orbits of each planet as non-interacting Keplerian orbits, and fit the light curve for best fit parameters of the mean stellar density, $\rho_{\star}$, the photometric zero point for each planet, the center of transit time for the first observed transit $T_{0}$, the orbital period $P$, the impact parameter $b$, the scaled planetary radius $R_{p}/R_{\star}$, and the components of the eccentricity vector, $e\cos\omega$ and $e\sin\omega$. To account for the TTVs, the light curve model cadence was contracted and expanded based on a linear interpolation of measured transit times for each planet to match the observed transit times. To calculate the posterior distributions of model paramters, we used an MCMC routine described in Section 4.1 of \citet{row13}. We used the determination of $e\cos\omega$ and $e\sin\omega$ from dynamical modeling of the TTVs as constraints characterized by a Gaussian distribution. We generated 4$\times$1,000,000 Markov-Chains and calculated the median value for each model parameter and its 1$\sigma$ uncertainty interval which we list in Table~\ref{tbl-transit}. 
\begin{table}[h!]
  \begin{center}
    \begin{tabular}{|cccccc|}
   \hline
Planet & $R_p/R_{\star}$                       &  depth (ppm)               & $b$                      & $i$ ($^{\circ}$)          & $a/R_{\star}$ \\ 
\hline
   b   &   0.02442$^{+0.00018}_{-0.00018}$   &  675.1$^{+8.3}_{-9.2}$   &  0.410$^{+0.021}_{-0.036}$ &  88.78$^{+0.07}_{-0.09}$    & 19.26$^{+0.19}_{-0.27}$  \\
   c   &   0.02618$^{+0.00019}_{-0.00016}$   &  789.9$^{+9.0}_{-9.5}$   &  0.278$^{+0.047}_{-0.037}$ &  89.48$^{+0.07}_{-0.09}$    & 30.90$^{+0.30}_{-0.43}$  \\
   d   &   0.05038$^{+0.00010}_{-0.00010}$   &  2968$^{+11}_{-13}$      &  0.056$^{+0.022}_{-0.056}$ &  89.93$^{+0.07}_{-0.03}$   & 47.42$^{+0.46}_{-0.67}$  \\
   e   &   0.02458$^{+0.00079}_{-0.00094}$   &  453$^{+26}_{-28}$      &  0.963$^{+0.015}_{-0.016}$ &   89.13$^{+0.02}_{-0.02}$   & 63.68$^{+0.62}_{-0.90}$   \\
       \hline
        \hline
    \end{tabular}
    \caption{Transit constraints on the planets of KOI-152, following dynamical models; $b$ signifies impact parameter, $i$ inclination of the orbit to to the plane of the sky and $a$ the orbital semimajor axis.}\label{tbl-transit}
  \end{center}
\end{table}

The light curve model gives a geometrical measurement of $\rho_{\star}$, which we combined with the spectroscopic determination of $T_{eff}$ and [Fe/H]. These we matched to stellar evolution models \citep{dem04}, to estimate the stellar mass and radius. Our uncertainties follow from the posterior distributions of our MCMC analysis. Our results are in Table~\ref{tbl-star}.
\begin{table}[h!]
  \begin{center}
    \begin{tabular}{||l|l||}
   \hline
\hline
$ M_{\star} (M_{\odot})$  &  1.165$^{+0.044}_{-0.045}$  \\
$ R_{\star} (R_{\odot})$  &  1.302$^{+0.026}_{-0.027}$  \\
$ L_{\star} (L_{\odot})$  &  2.20$^{+0.18}_{-0.22}$ \\
$T_{\rm eff}$ (K)    &  6174$^{+83}_{-117}$   \\
\logg\ (cm s$^{-2}$)    &  4.274$^{+0.012}_{-0.013}$  \\
$Z$  &   0.0169$^{+0.0026}_{-0.0030}$ \\
$\rho_{\star}$ (g cm$^{-3}$) & $0.741^{+0.026}_{-0.034}$  \\
Age (Gyr)  &   3.44$^{+0.60}_{-0.91}$ \\ 
       \hline
        \hline
    \end{tabular}
    \caption{The characteristics of the star KOI-152, with 1$\sigma$ uncertainties.}\label{tbl-star}
  \end{center}
\end{table}

\section{Non-transiting Perturbers?}
Near strong mean motion resonances, pairwise interactions cause a TTV frequency at the circulating argument of the nearest resonance. The period of this signal increases closer to the resonance. Thus in Table~\ref{tbl-resonances}, the highest values in the final column mark the orbit pairs closest to resonance. For the known candidates in this system, all pairwise TTVs due to first order resonance have a complete cycle of observed transit times, hence the well constrained model fit.
\begin{figure}
\includegraphics [height = 1.6 in]{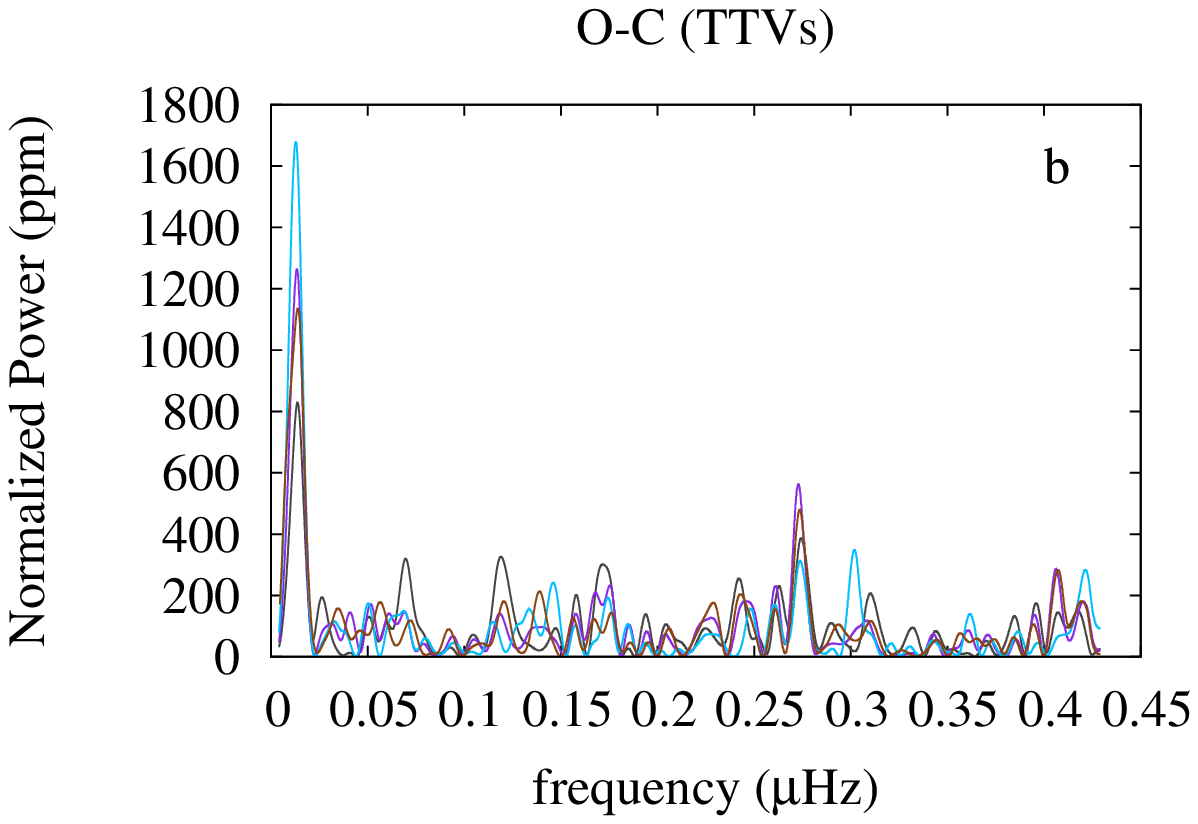}
\includegraphics [height = 1.6 in]{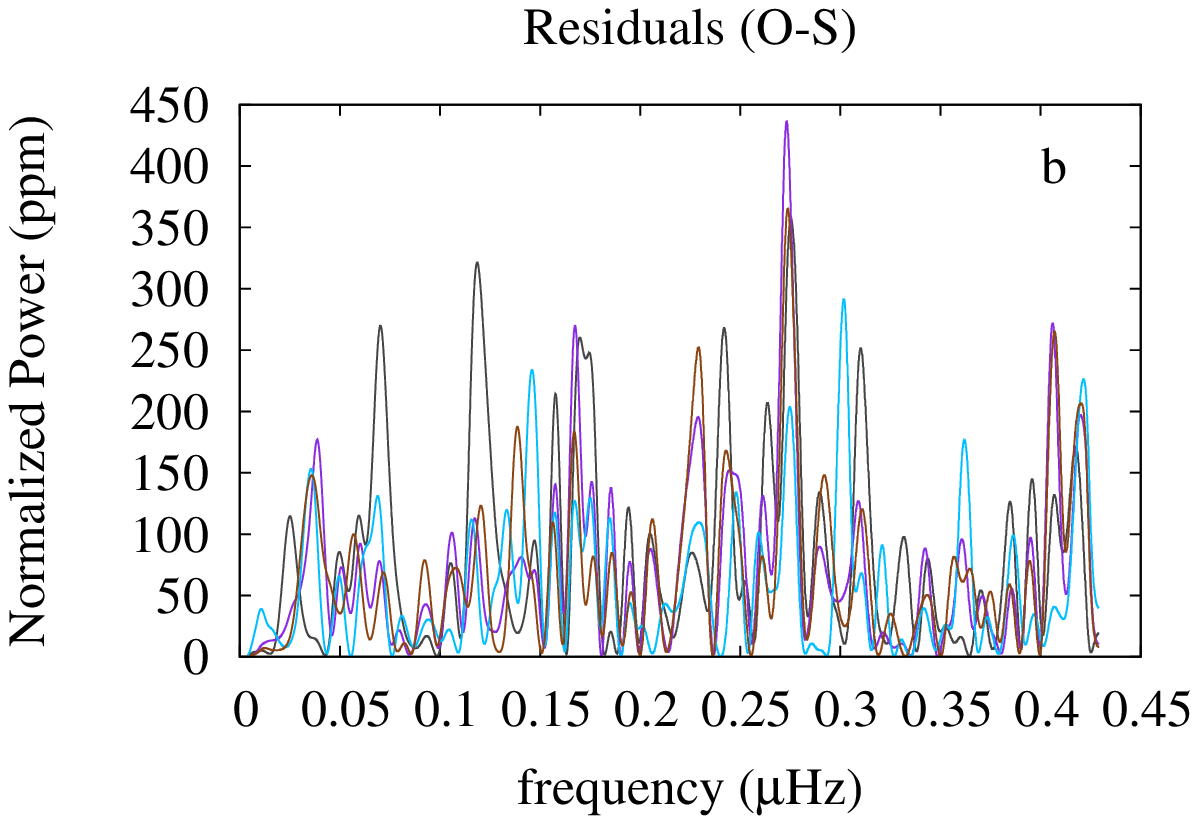}
\newline
\includegraphics [height = 1.6 in]{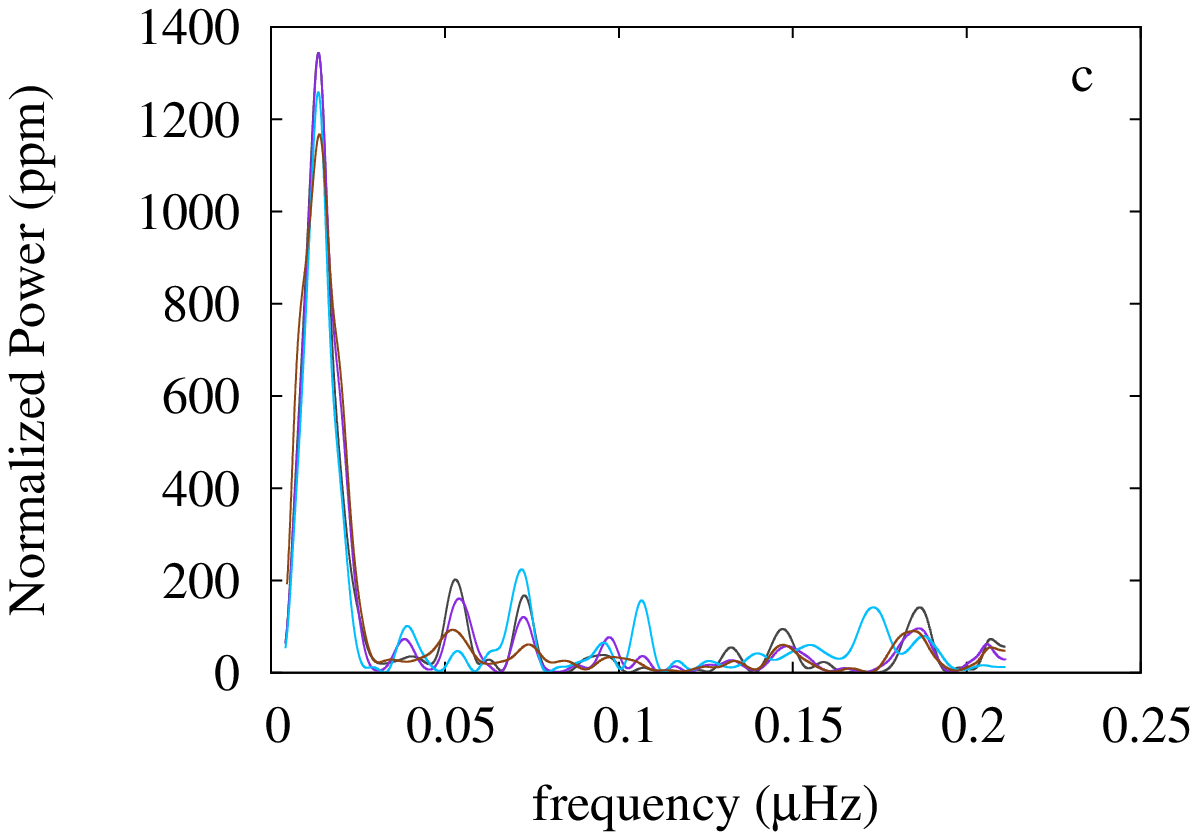}
\includegraphics [height = 1.6 in]{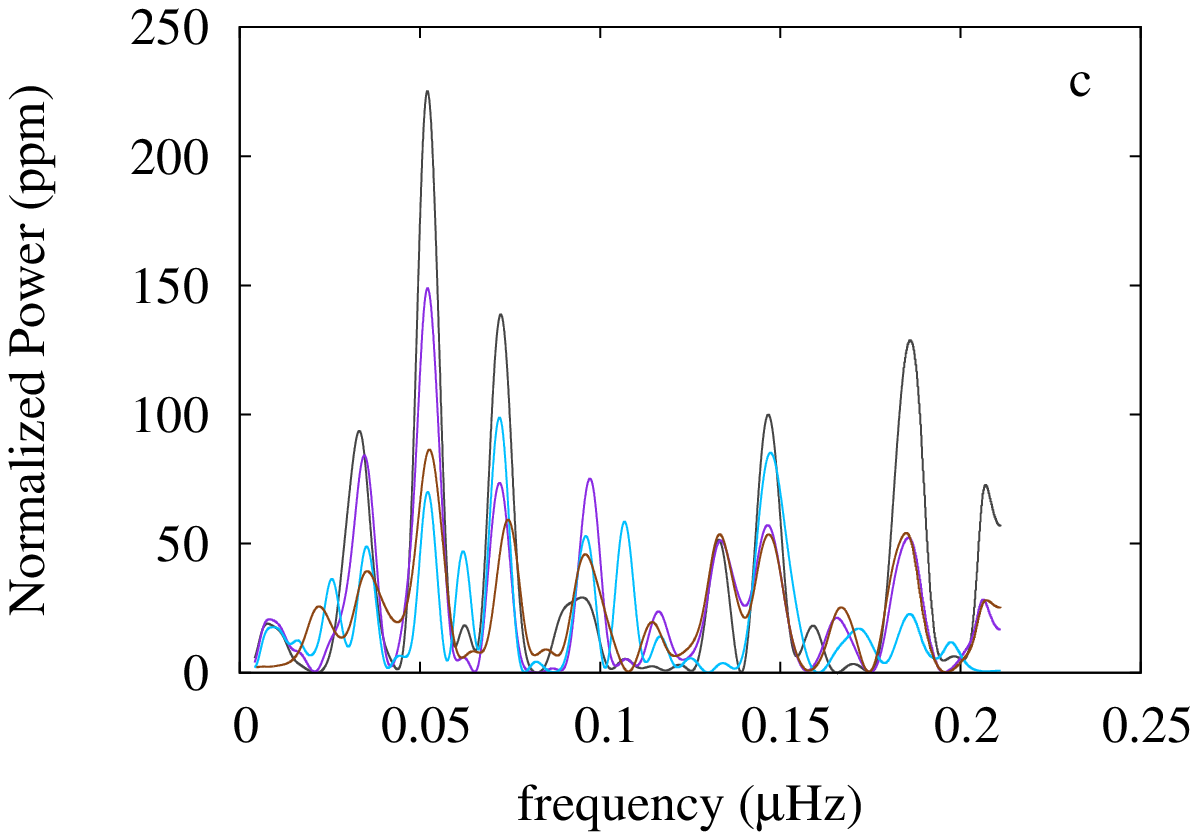}
\newline
\includegraphics [height = 1.6 in]{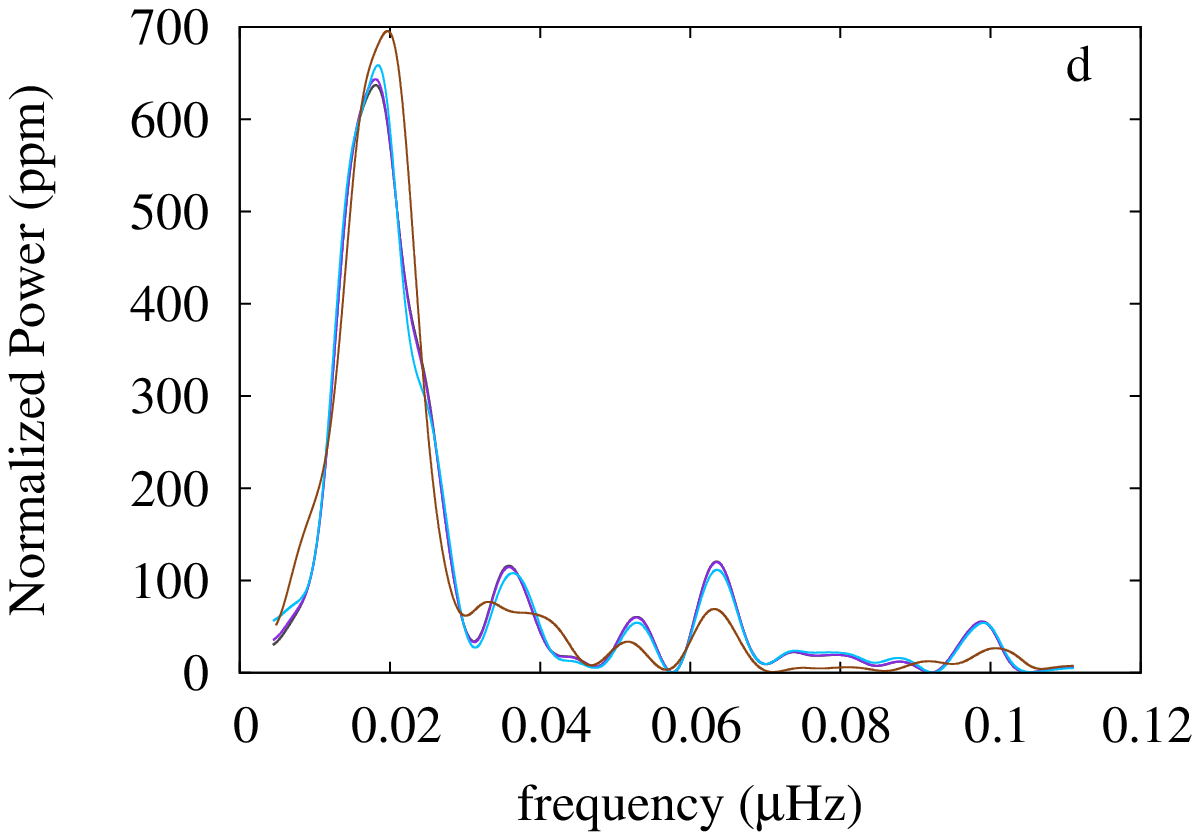}
\includegraphics [height = 1.6 in]{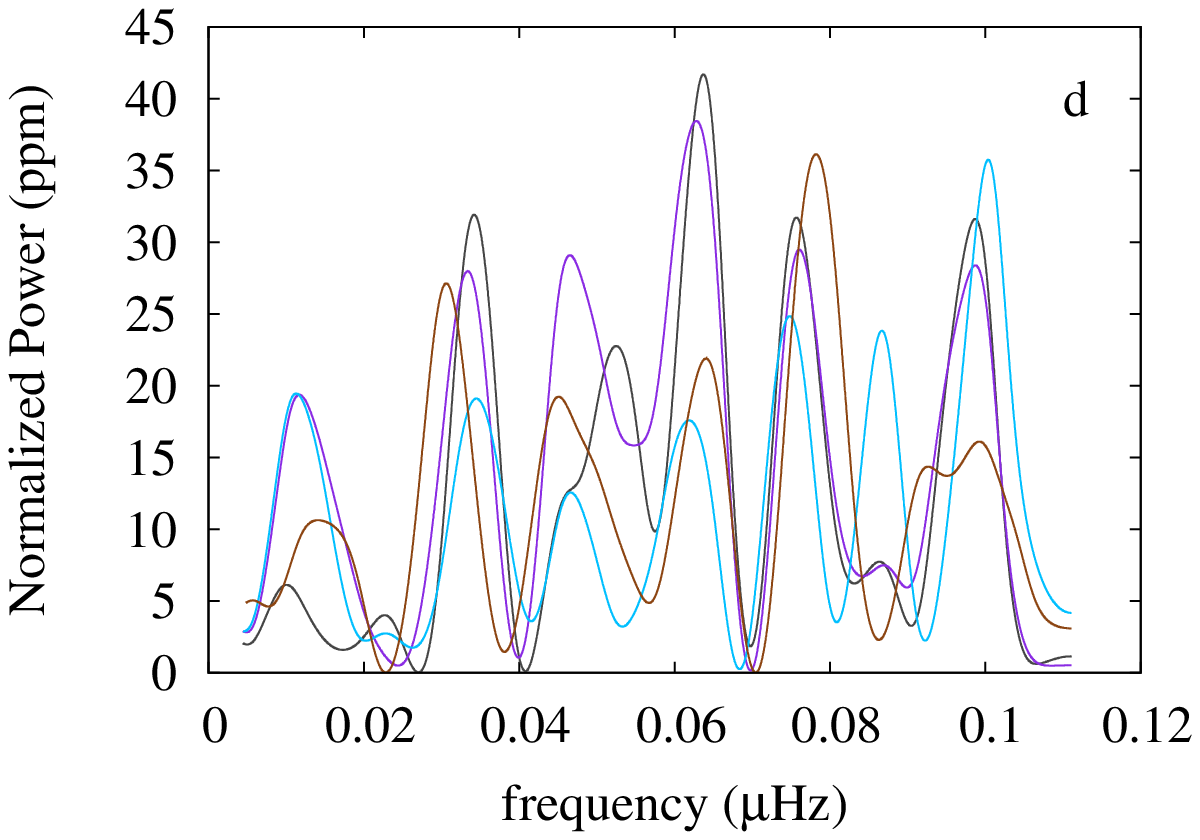}
\newline
\includegraphics [height = 1.6 in]{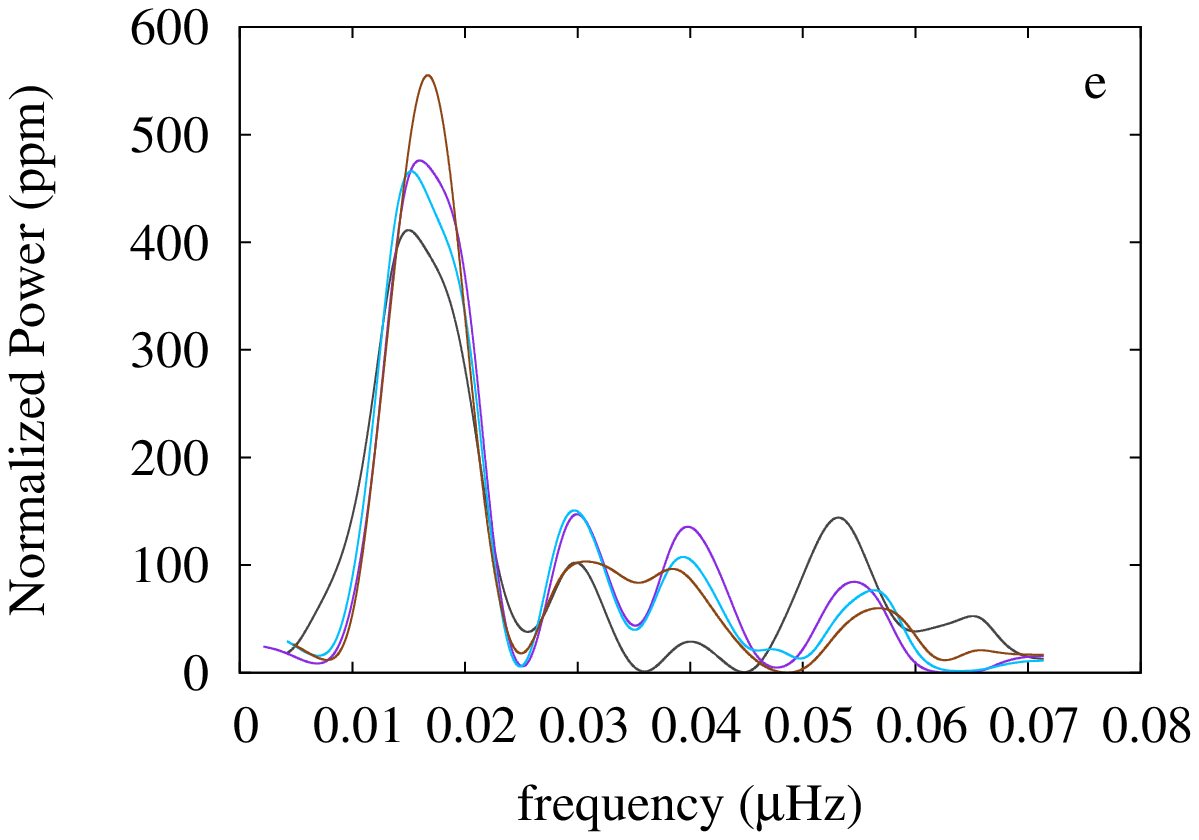}
\includegraphics [height = 1.6 in]{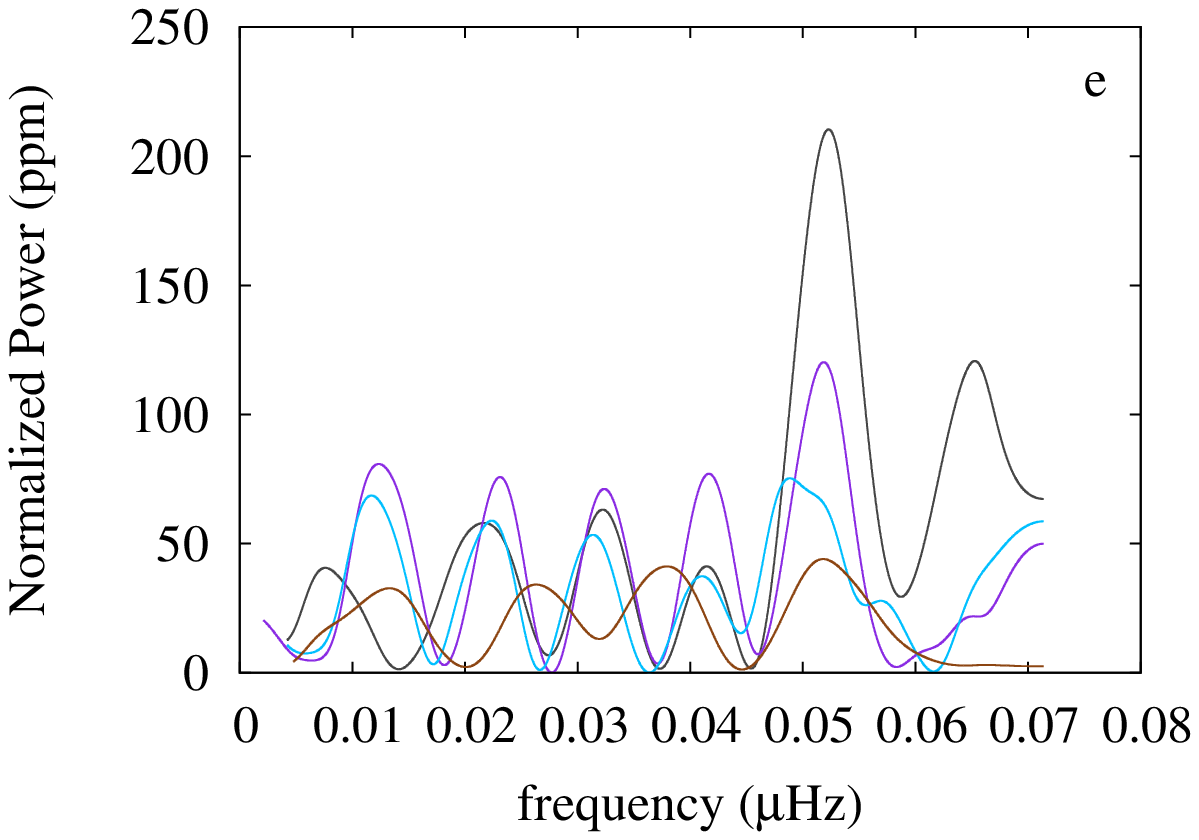}
\caption{Periodograms of the observed TTVs of the planets and residuals of best fit models for each of the four datasets of transit times. Here, the normalized power measured in parts per million (ppm), and the periodograms have been normalized by summing the power over 10,000 equally-spaced frequency channels between an assumed half-wave over the TTV baseline and the Nyquist frequency. The graphs highlight the dominant frequencies in the bestfit model (left) and the residuals (right). In each graph, the dark grey curves mark the best fit to the raw transit times, the purple curves mark the best fit to the combined SC and LC dataset with 3$\sigma$ outliers removed, the blue curves mark the best fit to the combined datset with all 2$\sigma$ outliers removed, and the brown curves mark the best fit to SC only times with 3$\sigma$ outliers removed.}
\label{fig:per-bcde}
\end{figure}

Figure~\ref{fig:per-bcde} displays the periodicities detected in the TTVs. For this figure we have constructed periodograms of the TTVs (O-C), as well as the residuals (O-S). For each periodogram the frequency was limited at the Nyquist frequency (half the sample rate as the maximum frequency, corresponding to twice the orbit period). The minimum frequency corresponds to the maximum period that is sampled over the observational baseline. For this limit, we have chosen twice the observational baseline, which would be an lower limit on any incomplete TTV period, or an upper limit on its frequency. In these plots, the absolute power in each peak need not correspond between models, since the combined datasets and the short cadence only dataset have different baselines and therefore different frequency domains. However, the relative height of peaks within each dataset, and their locations in frequency are of interest. In each case the peak in the periodogram corresponds to the expected TTV signal from the last column in Table~\ref{tbl-resonances}. However, the peaks are broad enough such that for the planets with two neighbors, `c' and `d', there are not two clear peaks in the periodogram. For `c', the single broad peak encompasses the expected TTV signal at 853 days (0.0136 $\mu$Hz) and 526 days (0.0220 $\mu$Hz). For `d' the shape of the broad curve hints at two peaks where we would expect them near 526 days (0.0220 $\mu$Hz) and 721 days (0.0161 $\mu$Hz) respectively, although they cannot be resolved.

There are no peaks in the residuals that stand out significantly from the background or noise. We expect more residual power at higher frequencies where outliers have been included in the fits, and this is confirmed in Fig.~\ref{fig:per-bcde}. We see no evidence that any unseen planets at KOI-152 contribute significantly to the TTVs of the four known candidates. Furthermore, it follows from the good fits that we have found to the transit time data at the expected TTV periods that these candidates are very likely to orbit the same star. Of course, we cannot rule out the possibility of other planets perturbing the four known candidates. Since the configuration is so compact, it appears that only an unseen perturber orbiting interior to KOI-152 b or beyond KOI-152 e could have any potential of confusing our four-planet model. We note that KOI-152 e has the highest fraction of its transits that are outliers, and that these are all in the second half of the data. Could this be due to the TTVs of an outer perturber increasing the TTV amplitude over the photometric baseline? Whilst we cannot exclude this possibility, we can consider how this hypothetical planet may affect our solutions. We focus on `e' since it is more likely to have an effect on our most surprising result, the low mass of KOI-152 d.

If a non-transiting planet had an orbital period near 104 days, its orbital period would be near 2:1 with KOI-152 d and 3:2 with KOI-152 e, and it could cause near-first-order resonance TTVs in KOI-152 d and e. Since we see no residual peaks in the periodogram for KOI-152 d, a perturber causing first-order TTVs on `d' appears unlikely. Hence, only if the TTVs induced on `d' had a similar period to the TTVs on `d' caused by `e' or `c' would our model for d's TTVs be confused. Any other possible near-first-order resonance with KOI-152 e would leave only high order TTVs on KOI-152 d, which would be an insignificant addition to its TTVs. Our solution to the mass of KOI-152 d is constrained by the well-fitted TTVs it induces on `c' and `e'. A false measurement of KOI-152 e's mass would cancel one but not both of these constraints, notwithstanding the freedom for eccentricities to adjust to a different mass for `e'. It appears unlikely that KOI-152 d, given its position between two transiting planets, would have a substantially different mass if a more distant planet were inducing TTVs in just one of its neighbors, namely, KOI-152 e. The effect of a fifth planet beyond KOI-152 e would have an even smaller effect on our solutions for KOI-152 b and c. 

 \section{Characterizing the Planets}
\begin{table}[h!] 
  \begin{center}
    \begin{tabular}{|ccccccc|}
      \hline
   \hline
      Planet  &  Mass ($M_{\oplus}$)     & Radius ($R_{\oplus}$)     &   Density (g cm$^{-3}$)       &  $a$  (AU)                            &   $e$    &     Flux ($F_{\odot, 1AU}$) \\
\hline
 b  & \textbf{10.9}$^{+7.4}_{-6.0}$     & \textbf{3.47}$^{+0.07}_{-0.07} $ & \textbf{1.43}$^{+0.97}_{-0.78}$  & \textbf{0.117}$^{+0.002}_{-0.002} $ &   \textbf{0.015}$^{+0.012}_{-0.006}$ &  \textbf{162} \\  
 c  & \textbf{5.9}$^{+1.9}_{-2.3}$     & \textbf{3.72}$^{+0.08}_{-0.08}$ & \textbf{0.62}$^{+0.20}_{-0.25}$   & \textbf{0.187}$^{+0.002}_{-0.003} $ &    \textbf{0.030}$^{+0.027}_{-0.021}$ &   \textbf{63} \\  
 d  & \textbf{6.0}$^{+2.1}_{-1.6}$     & \textbf{7.16}$^{+0.13}_{-0.16} $ & \textbf{0.09}$^{+0.03}_{-0.02}$  & \textbf{0.287}$^{+0.004}_{-0.004} $ &    \textbf{0.025}$^{+0.059}_{-0.023}$ &   \textbf{27} \\  
 e  & \textbf{4.1}$^{+1.2}_{-1.1}$     & \textbf{3.49}$^{+0.14}_{-0.14} $ & \textbf{0.53}$^{+0.15}_{-0.15}$  & \textbf{0.386}$^{+0.005}_{-0.005} $ &    \textbf{0.012}$^{+0.044}_{-0.005}$ &   \textbf{15} \\  
      \hline
        \hline
    \end{tabular}\label{tbl-planet}
    \caption{The planets of KOI-152. All uncertainties are 1$\sigma$ confidence intervals, with planetary masses and radii in Earth units, and density in g cm$^{-3}$. The flux (final column) is scaled to the flux received from the Sun at 1 AU.}
  \end{center}
\end{table}

Table \ref{tbl-planet} shows our measured masses, radii, densities and incidence fluxes for each planet based on our best fit solutions, with uncertainties extended to account for all four sets of transit times. The planetary bulk densities follow from the constraint in stellar density:
\begin{equation}
\rho_p = \left(\rho_{\star}\right)\left(\frac{M_p}{M_{\star}}\right)\left(\frac{R_p}{R_{\star}}\right)^{-3}.
\label{rhoplanet}
\end{equation}
Each of the terms in parentheses in Equation~\ref{rhoplanet} is an independent source of uncertainty that we add in quadrature. We use this formula because it provides a more appropriate accounting of the uncertainties than the uncertainties in the planetary masses and radii. Due to the tight constraints on stellar and planetary radii, the dominant source of uncertainty here is planet-to-star mass ratios from our dynamical models. Nevertheless, all four candidates can be characterized as having low bulk density, due to the retention of volatiles. KOI-152 b has the highest bulk density (albeit with large uncertainties), despite having a similar radius to KOI-152 c and e. The nominal density of KOI-152 b is slightly less than a pure ice world of the same mass, suggesting either a substantial mass fraction of water and/or a relatively thin H/He envelope, with heavy elements in the interior. KOI-152 c and e have similar characteristics to the planets of Kepler-11 d and e (\citet{liss13}), and likely have envelopes that are far more significant by volume than by mass \citep{liss13}. The bulk density of KOI-152 d is the lowest for a \textit{Kepler} planet measured to date. Although the temperature of KOI-152 d is unknown, its place on the mass-radius diagram appears to give its envelope a total mass (as a fraction of the planetary mass) of around 50$\%$ at T = 500K, roughly 10$\%$ at 1000 K, \citep{rog11}. The equilibrium temperature of a planet assumes zero albedo and no internal heat source, and depends solely on the temperature of the host star and the orbital distance. For low eccentricities, 
\begin{equation}
T_{eq} \approx T_{\star}\sqrt{\frac{R_{\star}}{2a}}.
\label{Tbb}
\end{equation}
 Given its distance from the star, the equilibrium temperature of KOI-152 d is 634$\pm$ 16 K. Hence, it appears reasonable that KOI-152 d has an H/He envelope that contributes significantly more than 10$\%$ of the planetary mass, but less than 50$\%$ of the mass.

\section{KOI-152's Planets on the Mass-Radius Diagram}
Here we plot neptunes and sub-neptunes on the mass-radius diagram, including all planets with measured radii and masses, with nominal mass determinations up to 30 $M_{\oplus}$. These include mass determinations by radial velocity spectroscopy (RV) as well as TTVs. We adopt the data from published studies that include the most recent stellar spectral classification and mass determinations for: HAT-P-11 b (Kepler-3 b) \citep{bak10} HAT-P-26 b \citep{har11}, 55 Cancri e (\citealt{von11,win11,gill12}), GJ 3470 b \citep{bon12}, GJ 436 b (\citealt{ehr11,bal10}), GJ 1214 b (\citealt{cha09,val13}), CoRoT-7 b (\citealt{fer11,bru10}), Kepler-4 b, \citep{bor10}, Kepler-10 b \citep{bat11}, Kepler-11 b-f \citep{liss13}, Kepler-18 b-d \citep{coch11}, Kepler-30 b and d \citep{san12},  Kepler-36 b and c \citep{car12}, Kepler-68 b \citep{gil13}, and KOI-94b \citep{weis13}. We exclude KOI-94 c  ($M_p = 15.6^{+5.7}_{-15.6} M_{\oplus}$) and KOI-94 e ($M_p = 35^{+18}_{-28} M_{\oplus}$) from the mass-radius diagram because their masses are poorly constrained, and in the case of KOI-94 e, its nominal mass is beyond 30 $M_{\oplus}$. For the solar system, we adopt data given in \citet{dpl06}.

\begin{figure}[h!]
\includegraphics [height = 2.9 in]{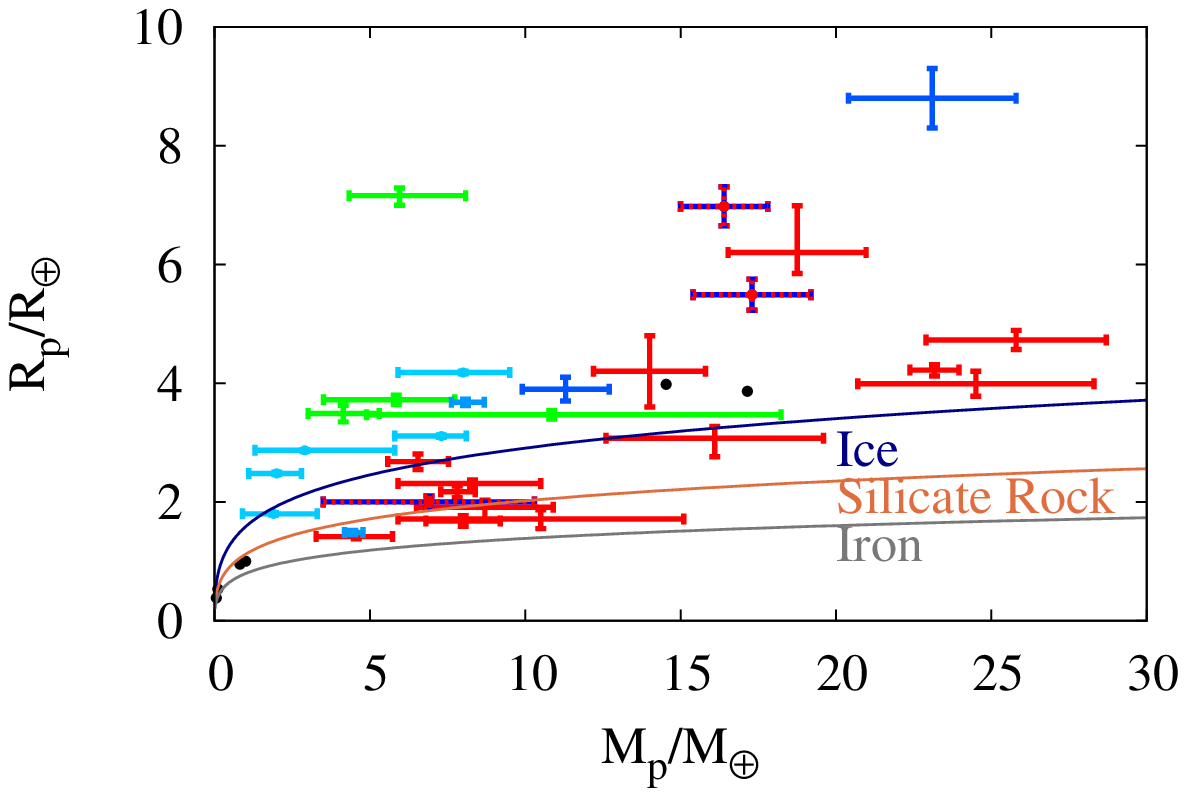}
\newline
\includegraphics [height = 2.9 in]{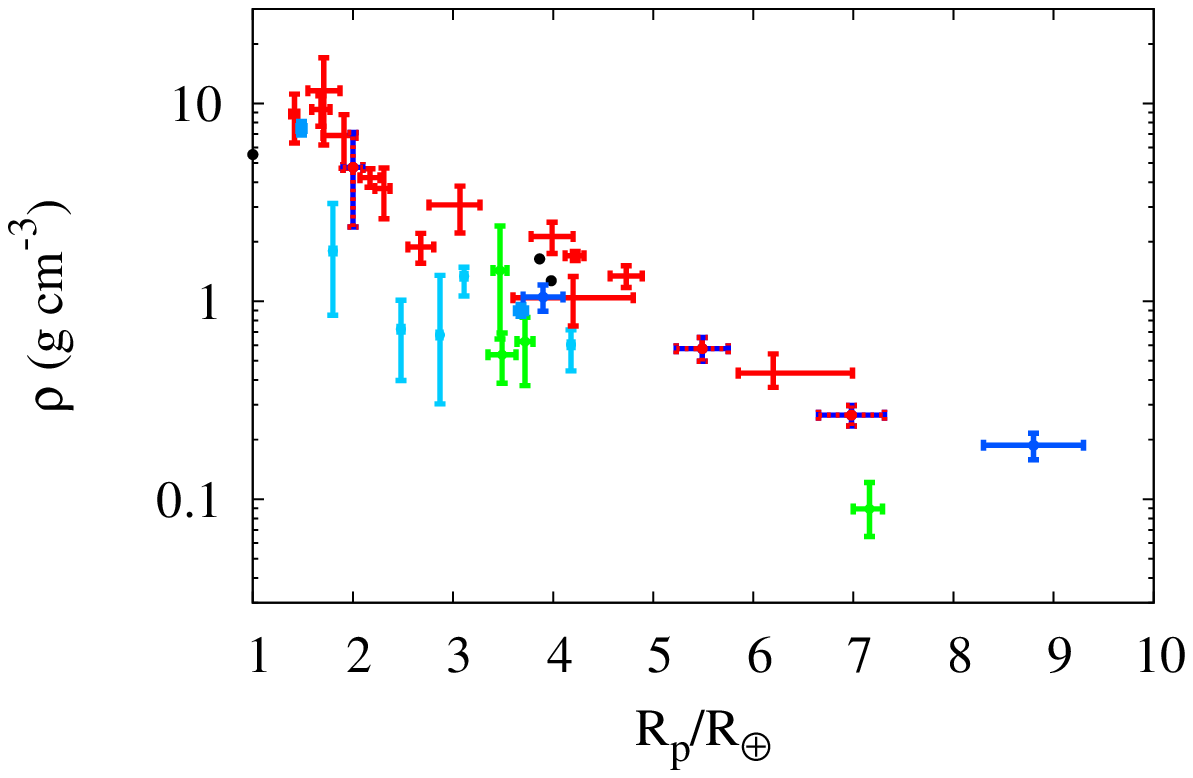}
\caption{The measured masses and radii of all known planets with masses $<30$ $M_{\oplus}$. Planets of the solar system are depicted by black points, transiting planets with masses measured by radial velocity spectroscopy (RV) are in red, planets with masses measured by TTVs are in blue, and the planets of KOI-152 are in green. The masses of the planets orbiting Kepler-18 are the combined result of RV and TTV observations, and appear as red and blue points above. The curves in the left panel represent solutions for the radii of planets composed of pure water ice, silicate rock, or iron \citep{for07}. In the bottom panel, we plot bulk densities as a function of planet size.}
\label{fig:M-R} 
\end{figure}

The planets in Fig.~\ref{fig:M-R} are all compared to theoretical models of pure water ice, silicate rock, or iron planets. Model curves for planetary radii of planets made from pure ice, rock or iron follow the results of \citet{for07}. Several features stand out from Fig.~\ref{fig:M-R}. Firstly, amongst the sub-neptunes, TTV planets are systematically larger and hence less dense than RV planets in the same mass range. This is most likely due to the biases of both techniques. Since RV detectability declines rapidly with orbital distance, few RV masses have been measured beyond a few days orbital period, and several of the planets in the RV sample may well have suffered mass loss from the intense stellar radiation upon their atmospheres. On the other hand, the TTV systems are all dynamically compact multiplanet systems, where for a given planetary size, smaller masses are more likely to permit stable orbits (even though a larger mass makes the TTVs more detectable). Furthermore, for planets in multiplanets systems of a particular mass, the larger (and hence lower density) planets can have their transit times measured more accurately for TTV anlysis.
The bottom panel of Fig.~\ref{fig:M-R} highlights the two orders of magnitude in the range in planetary densities that are observed in the mass range up to 30 $M_{\oplus}$. The RV density determinations appear more tightly correlated than the TTV determinations, although the uncertainty in density for the low mass, low density TTV planets are certainly larger. Note that in this plot we show planetary radii from a mass limited sample. For given masses, and thus given detectability considerations, larger planets have more precise transit timing measurements. There is thus a bias in TTV detections to large, low density planets, and not massive, smaller planets. This selection effect is part of the reason that the $R_p-\rho_p$ curve declines to the right. Nominally, KOI-152 d has the lowest planetary bulk density of exoplanets measured to date, although Kepler-12 b, with a mass of 137 $M_{\oplus}$, and radius of 19 $R_{\oplus}$, has a very similar bulk density within uncertainties. KOI-152 d is twenty times less massive than Kepler-12 b, and also receives roughly 30 times less insolation than Kepler-12 b \citep{for11}. It is significantly less dense than other characterized sub-Saturn mass planets. 

\begin{figure}[h!]
\includegraphics [height = 2.9 in]{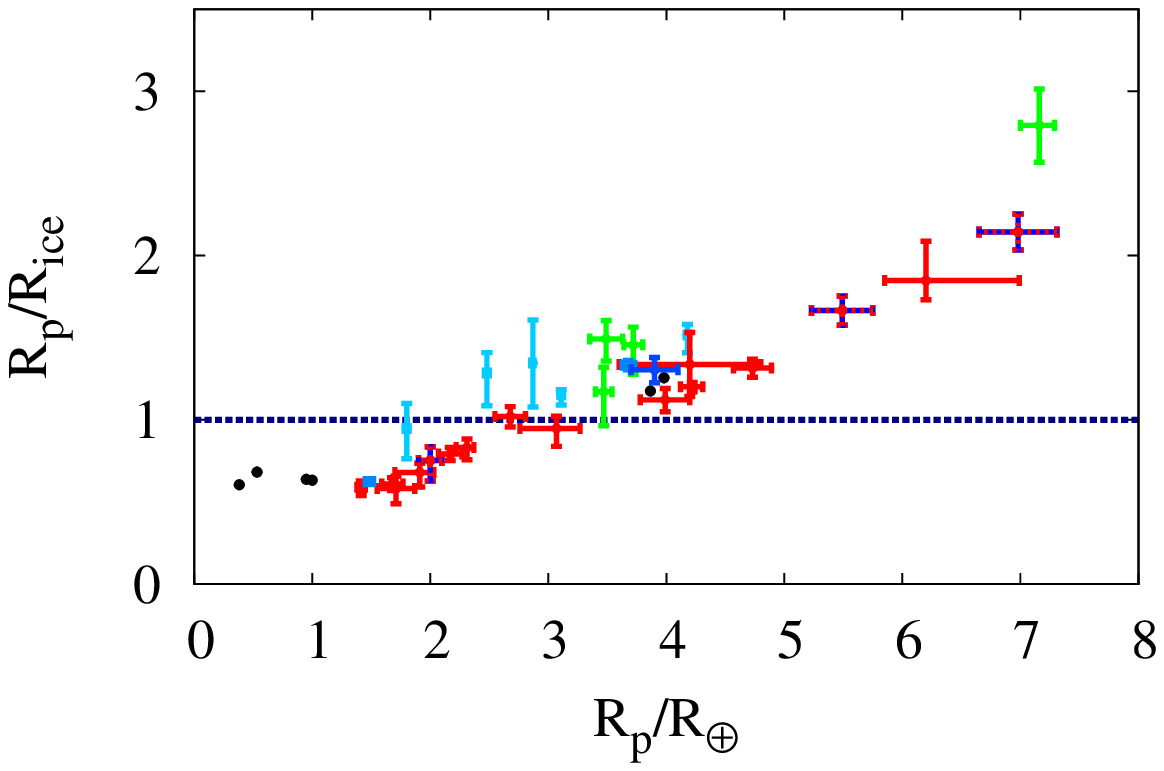}
\newline
\includegraphics [height = 2.9 in]{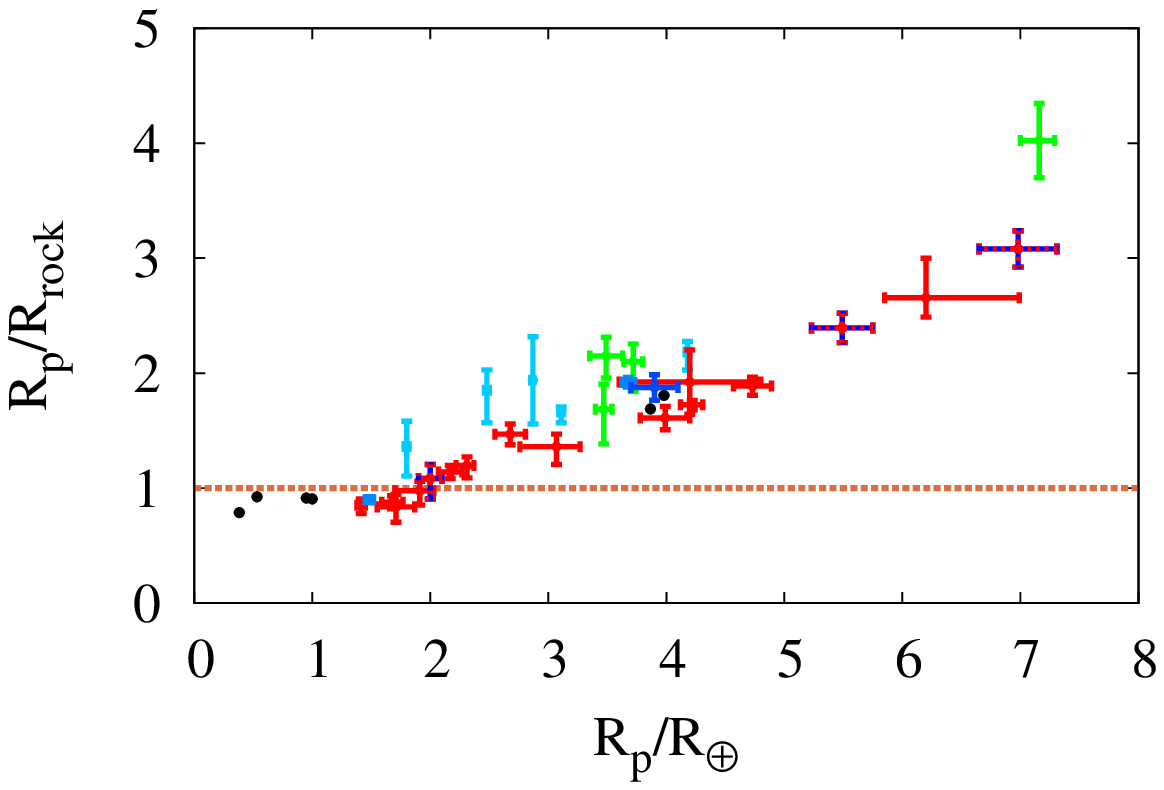}
\caption{Planetary radii compared to solutions of pure ice (top panel) or pure silicate rock (bottom panel). Mass determinations by RV are in red, and TTV in blue, with the planets of KOI-152 in green. Planets above the dotted blue line in the left panel, are less dense than water ice and are likely to have deep H/He envelope. Planets below the dotted brown line in the bottom panel are denser than rock, and are unlikely to have a substantial atmosphere. Planets between these limits must contain some volatiles, but the unknown tradeoff between water or a combination of H/He and rock precludes a definitive answer on whether there is an H/He envelope.}
\label{fig:iceline} 
\end{figure}
If we parametrize the ratio of the planetary radius to the radius of a planet of the same mass composed purely of ice, or rock, we have a simple test of whether the planet has retained a substantial gaseous envelope. Planets that are less dense than water ice, (above the blue line in the top panel of Fig.~\ref{fig:iceline}), are likely to have a volumetrically substantial H/He envelope. Planets that are more dense than pure silicate rock, (below the brown line in the bottom panel of Fig.~\ref{fig:iceline}), are less likely to retain a substantial amount of volatiles. Between these two limits, the atmosphere of a planet cannot be determined without more information, since it is unknown in what proportions volatiles are likely to be present in the form of ices and gases. Nevertheless, we surmise from Fig.~\ref{fig:iceline} that planets with $R_{p} < 2R_{\oplus}$ appear unlikely to retain significant atmosphere, and from the bottom panel that planets with $R_{p} > 4R_{\oplus}$ are very likely to retain deep atmospheres. All the characterized planets in our chosen mass range show remarkably little scatter in Fig.~\ref{fig:iceline}.


\begin{figure}[h!]
\includegraphics [height = 3.1 in]{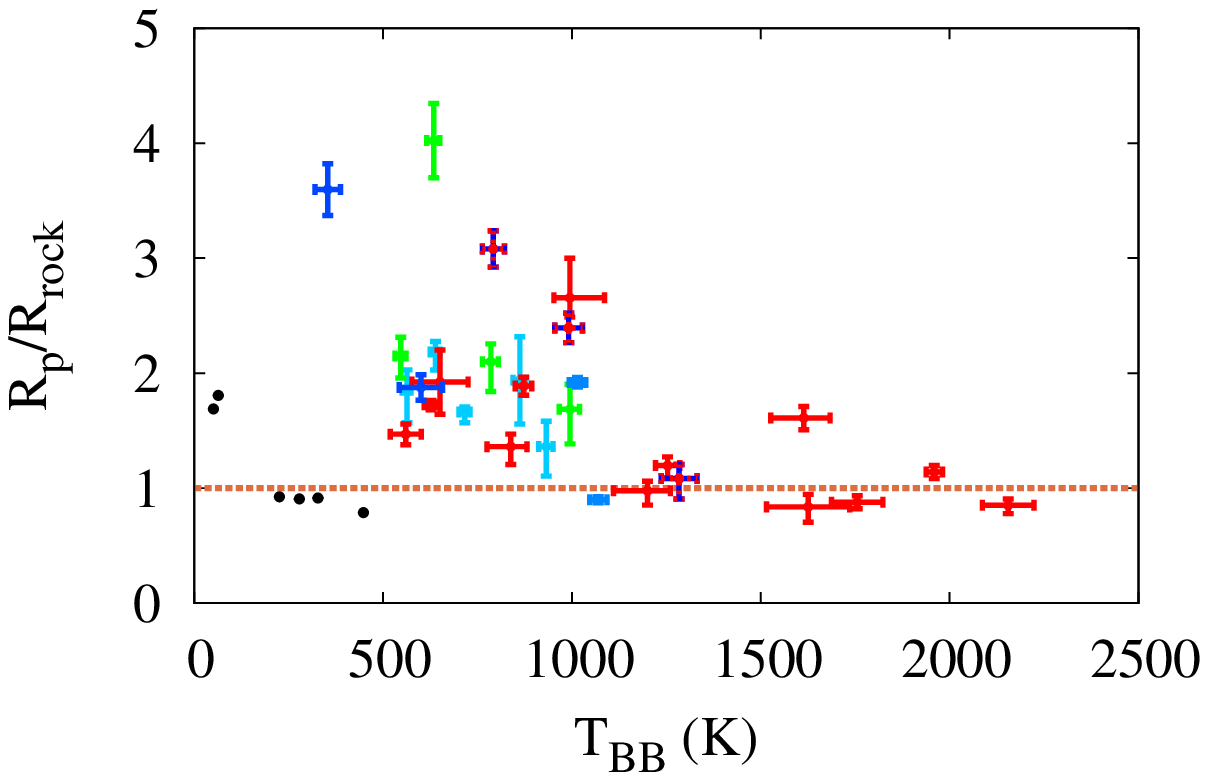}
\caption{Planetary radii compared to solutions of pure rock (left panel) as a function of equilibrium temperature. Mass determinations by RV are in red, and TTV in blue, with the planets of KOI-152 in green. A sharp transition to planets with little or no volatiles appears around 1100K. In the cooler regime, no planets that are denser than rock are seen amongst the exoplanets.}
\label{fig:RFrockline} 
\end{figure}
In Fig.~\ref{fig:RFrockline}, we show the equilibrium temperature of the planet against their radii relative to pure rock or ice. We see an abrupt transition around 1050 K. At higher temperatures, there is no evidence of planets harboring substantial gases, and below this temperature, a wide range of densities appears permissible. One possible exception is Kepler-4b, with a equilibrium temperature of 1614 K, $R_p/R_{rock} =1.61$, and $R_p/R_{ice} = 1.12$. \citet{bor10} estimated that Kepler-4b retains a deep H/He atmosphere of about 4-6$\%$ by mass. Of the ``hot" planets below 30 $M_{\oplus}$, Kepler-4b is alone in requiring a deep atmosphere. Nevertheless, we still see a much wider range in bulk densities amongst the cool sub-neptunes. Kepler-36 stands out as having two planets that are close in orbital periods (13.84 and 16.24 days respectively, \citealt{car12}), with remarkably different densities. These planets appear to straddle the transition in equilibrium temperature at $\sim 1050$ K, with the outer planet retaining an atmosphere, and the inner one denser than pure silicate rock \citep{lop13}. KOI-152 b lies close to Kepler-36 c on this plot, with a blackbody temperature of 1004 $\pm$ 24 K, although it is marginally denser. Amongst the ``cool'' sub-neptunes, no exoplanets that are denser than pure rock have yet been discovered.
\section{Conclusion}
Assuming a four-planet model for the TTVs in the four transiting candidates of KOI-152, we have added four sub-Uranus masses to the mass-radius diagram, bringing the total to twenty exoplanets. We confirm the planetary nature of these candidates and that they are planets in the same system. The planets of KOI-152 appear to follow the trend of the planets orbiting Kepler-11, being of low density with significant envelopes. Nevertheless, the planets show remarkable variety in their bulk masses. KOI-152 b, c and e all have radii between 3.5 and 4.0 $R_{\oplus}$, and yet their masses range from $\sim$3 to $\sim$13 $M_{\oplus}$, and their bulk densities range from 0.3 to 1.6 g cm$^{-3}$. The largest planet, KOI-152 d, ($\sim 7R_{\oplus}$), has a remarkably low mass given its size, and most likely has the lowest nominal bulk density amongst known planets with sub-Saturn masses.

D.J. gratefully acknowledges the support of the NASA Postdoctoral Program.


\begin{thebibliography}{27}
\expandafter\ifx\csname natexlab\endcsname\relax\def\natexlab#1{#1}\fi


\bibitem[{{Bakos} {et~al.}(2010){Bakos}, {Torres}, {P{\'a}l}, {Hartman},
  {Kov{\'a}cs}, {Noyes}, {Latham}, {Sasselov}, {Sip{\H o}cz}, {Esquerdo},
  {Fischer}, {Johnson}, {Marcy}, {Butler}, {Isaacson}, {Howard}, {Vogt},
  {Kov{\'a}cs}, {Fernandez}, {Mo{\'o}r}, {Stefanik}, {L{\'a}z{\'a}r}, {Papp},
  \& {S{\'a}ri}}]{bak10}
{Bakos}, G.~{\'A}., {et~al.} 2010, \apj, 710, 1724

\bibitem[{{Ballard} {et~al.}(2010){Ballard}, {Charbonneau}, {Deming},
  {Knutson}, {Christiansen}, {Holman}, {Fabrycky}, {Seager}, \&
  {A'Hearn}}]{bal10}
{Ballard}, S., {et~al.} 2010, \pasp, 122, 1341

\bibitem[{{Batalha} {et~al.}(2011){Batalha}, {Borucki}, {Bryson}, {Buchhave},
  {Caldwell}, {Christensen-Dalsgaard}, {Ciardi}, {Dunham}, {Fressin},
  {Gautier}, {Gilliland}, {Haas}, {Howell}, {Jenkins}, {Kjeldsen}, {Koch},
  {Latham}, {Lissauer}, {Marcy}, {Rowe}, {Sasselov}, {Seager}, {Steffen},
  {Torres}, {Basri}, {Brown}, {Charbonneau}, {Christiansen}, {Clarke},
  {Cochran}, {Dupree}, {Fabrycky}, {Fischer}, {Ford}, {Fortney}, {Girouard},
  {Holman}, {Johnson}, {Isaacson}, {Klaus}, {Machalek}, {Moorehead},
  {Morehead}, {Ragozzine}, {Tenenbaum}, {Twicken}, {Quinn}, {VanCleve},
  {Walkowicz}, {Welsh}, {Devore}, \& {Gould}}]{bat11}
{Batalha}, N.~M., {et~al.} 2011, \apj, 729, 27

\bibitem[{{Bonfils} {et~al.}(2012){Bonfils}, {Gillon}, {Udry}, {Armstrong},
  {Bouchy}, {Delfosse}, {Forveille}, {Fumel}, {Jehin}, {Lendl}, {Lovis},
  {Mayor}, {McCormac}, {Neves}, {Pepe}, {Perrier}, {Pollaco}, {Queloz}, \&
  {Santos}}]{bon12}
{Bonfils}, X., {et~al.} 2012, \aap, 546, A27

\bibitem[{{Borucki} {et~al.}(2010){Borucki}, {Koch}, {Brown}, {Basri},
  {Batalha}, {Caldwell}, {Cochran}, {Dunham}, {Gautier}, {Geary}, {Gilliland},
  {Howell}, {Jenkins}, {Latham}, {Lissauer}, {Marcy}, {Monet}, {Rowe}, \&
  {Sasselov}}]{bor10}
{Borucki}, W.~J., {et~al.} 2010, \apjl, 713, L126

\bibitem[{{Bruntt} {et~al.}(2010){Bruntt}, {Deleuil}, {Fridlund}, {Alonso},
  {Bouchy}, {Hatzes}, {Mayor}, {Moutou}, \& {Queloz}}]{bru10}
{Bruntt}, H., {et~al.} 2010, \aap, 519, A51

\bibitem[{{Carter} {et~al.}(2012){Carter}, {Agol}, {Chaplin}, {Basu},
  {Bedding}, {Buchhave}, {Christensen-Dalsgaard}, {Deck}, {Elsworth},
  {Fabrycky}, {Ford}, {Fortney}, {Hale}, {Handberg}, {Hekker}, {Holman},
  {Huber}, {Karoff}, {Kawaler}, {Kjeldsen}, {Lissauer}, {Lopez}, {Lund},
  {Lundkvist}, {Metcalfe}, {Miglio}, {Rogers}, {Stello}, {Borucki}, {Bryson},
  {Christiansen}, {Cochran}, {Geary}, {Gilliland}, {Haas}, {Hall}, {Howard},
  {Jenkins}, {Klaus}, {Koch}, {Latham}, {MacQueen}, {Sasselov}, {Steffen},
  {Twicken}, \& {Winn}}]{car12}
{Carter}, J.~A., {et~al.} 2012, Science, 337, 556

\bibitem[{{Charbonneau} {et~al.}(2009){Charbonneau}, {Berta}, {Irwin}, {Burke},
  {Nutzman}, {Buchhave}, {Lovis}, {Bonfils}, {Latham}, {Udry}, {Murray-Clay},
  {Holman}, {Falco}, {Winn}, {Queloz}, {Pepe}, {Mayor}, {Delfosse}, \&
  {Forveille}}]{cha09}
{Charbonneau}, D., {et~al.} 2009, \nat, 462, 891

\bibitem[{{Claret} \& {Bloemen}(2011)}]{cla11}
{Claret}, A., \& {Bloemen}, S. 2011, \aap, 529, A75

\bibitem[{{Cochran} {et~al.}(2011){Cochran}, {Fabrycky}, {Torres}, {Fressin},
  {D{\'e}sert}, {Ragozzine}, {Sasselov}, {Fortney}, {Rowe}, {Brugamyer},
  {Bryson}, {Carter}, {Ciardi}, {Howell}, {Steffen}, {Borucki}, {Koch}, {Winn},
  {Welsh}, {Uddin}, {Tenenbaum}, {Still}, {Seager}, {Quinn}, {Mullally},
  {Miller}, {Marcy}, {MacQueen}, {Lucas}, {Lissauer}, {Latham}, {Knutson},
  {Kinemuchi}, {Johnson}, {Jenkins}, {Isaacson}, {Howard}, {Horch}, {Holman},
  {Henze}, {Haas}, {Gilliland}, {Gautier}, {Ford}, {Fischer}, {Everett},
  {Endl}, {Demory}, {Deming}, {Charbonneau}, {Caldwell}, {Buchhave}, {Brown},
  \& {Batalha}}]{coch11}
{Cochran}, W.~D., {et~al.} 2011, \apjs, 197, 7

\bibitem[{{Demarque} {et~al.}(2004){Demarque}, {Woo}, {Kim}, \& {Yi}}]{dem04}
{Demarque}, P., {Woo}, J.-H., {Kim}, Y.-C., \& {Yi}, S.~K. 2004, \apjs, 155,
  667

\bibitem[{{de Pater} \& {Lissauer}(2010)}]{dpl06}
{de Pater}, I., \& {Lissauer}, J. 2010, {Planetary Sciences} (Cambridge
  University Press)

\bibitem[{{Ehrenreich} {et~al.}(2011){Ehrenreich}, {Lecavelier Des Etangs}, \&
  {Delfosse}}]{ehr11}
{Ehrenreich}, D., {Lecavelier Des Etangs}, A., \& {Delfosse}, X. 2011, \aap,
  529, A80

\bibitem[{{Ferraz-Mello} {et~al.}(2011){Ferraz-Mello}, {Tadeu Dos Santos},
  {Beaug{\'e}}, {Michtchenko}, \& {Rodr{\'{\i}}guez}}]{fer11}
{Ferraz-Mello}, S., {Tadeu Dos Santos}, M., {Beaug{\'e}}, C., {Michtchenko},
  T.~A., \& {Rodr{\'{\i}}guez}, A. 2011, \aap, 531, A161

\bibitem[{{Gautier} {et~al.}(2012){Gautier}, {Charbonneau}, {Rowe}, {Marcy},
  {Isaacson}, {Torres}, {Fressin}, {Rogers}, {D{\'e}sert}, {Buchhave},
  {Latham}, {Quinn}, {Ciardi}, {Fabrycky}, {Ford}, {Gilliland}, {Walkowicz},
  {Bryson}, {Cochran}, {Endl}, {Fischer}, {Howell}, {Horch}, {Barclay},
  {Batalha}, {Borucki}, {Christiansen}, {Geary}, {Henze}, {Holman}, {Ibrahim},
  {Jenkins}, {Kinemuchi}, {Koch}, {Lissauer}, {Sanderfer}, {Sasselov},
  {Seager}, {Silverio}, {Smith}, {Still}, {Stumpe}, {Tenenbaum}, \& {Van
  Cleve}}]{gau12}
{Gautier}, III, T.~N., {et~al.} 2012, \apj, 749, 15

\bibitem[{{Fortney} {et~al.}(2007){Fortney}, {Marley}, \& {Barnes}}]{for07}
{Fortney}, J.~J., {Marley}, M.~S., \& {Barnes}, J.~W. 2007, \apj, 659, 1661

\bibitem[{{Fortney} {et~al.}(2011){Fortney}, {Demory}, {D{\'e}sert}, {Rowe},
  {Marcy}, {Isaacson}, {Buchhave}, {Ciardi}, {Gautier}, {Batalha}, {Caldwell},
  {Bryson}, {Nutzman}, {Jenkins}, {Howard}, {Charbonneau}, {Knutson}, {Howell},
  {Everett}, {Fressin}, {Deming}, {Borucki}, {Brown}, {Ford}, {Gilliland},
  {Latham}, {Miller}, {Seager}, {Fischer}, {Koch}, {Lissauer}, {Haas}, {Still},
  {Lucas}, {Gillon}, {Christiansen}, \& {Geary}}]{for11}
{Fortney}, J.~J., {et~al.} 2011, \apjs, 197, 9

\bibitem[{{Gilliland} {et~al.}(2013){Gilliland}, {Marcy}, {Rowe}, {Rogers},
  {Torres}, {Fressin}, {Lopez}, {Buchhave}, {Christensen-Dalsgaard},
  {D{\'e}sert}, {Henze}, {Isaacson}, {Jenkins}, {Lissauer}, {Chaplin}, {Basu},
  {Metcalfe}, {Elsworth}, {Handberg}, {Hekker}, {Huber}, {Karoff}, {Kjeldsen},
  {Lund}, {Lundkvist}, {Miglio}, {Charbonneau}, {Ford}, {Fortney}, {Haas},
  {Howard}, {Howell}, {Ragozzine}, \& {Thompson}}]{gil13}
{Gilliland}, R.~L., {et~al.} 2013, \apj, 766, 40

\bibitem[{{Gillon} {et~al.}(2012){Gillon}, {Demory}, {Benneke}, {Valencia},
  {Deming}, {Seager}, {Lovis}, {Mayor}, {Pepe}, {Queloz}, {S{\'e}gransan}, \&
  {Udry}}]{gill12}
{Gillon}, M., {et~al.} 2012, \aap, 539, A28

\bibitem[{{Hartman} {et~al.}(2011){Hartman}, {Bakos}, {Kipping}, {Torres},
  {Kov{\'a}cs}, {Noyes}, {Latham}, {Howard}, {Fischer}, {Johnson}, {Marcy},
  {Isaacson}, {Quinn}, {Buchhave}, {B{\'e}ky}, {Sasselov}, {Stefanik},
  {Esquerdo}, {Everett}, {Perumpilly}, {L{\'a}z{\'a}r}, {Papp}, \&
  {S{\'a}ri}}]{har11}
{Hartman}, J.~D., {et~al.} 2011, \apj, 728, 138

\bibitem[{{Huber} {et~al.}(2013){Huber}, {Chaplin}, {Christensen-Dalsgaard},
  {Gilliland}, {Kjeldsen}, {Buchhave}, {Fischer}, {Lissauer}, {Rowe},
  {Sanchis-Ojeda}, {Basu}, {Handberg}, {Hekker}, {Howard}, {Isaacson},
  {Karoff}, {Latham}, {Lund}, {Lundkvist}, {Marcy}, {Miglio}, {Silva Aguirre},
  {Stello}, {Arentoft}, {Barclay}, {Bedding}, {Burke}, {Christiansen},
  {Elsworth}, {Haas}, {Kawaler}, {Metcalfe}, {Mullally}, \& {Thompson}}]{hub13}
{Huber}, D., {et~al.} 2013, \apj, 767, 127

\bibitem[{{Lissauer} {et~al.}(2011){Lissauer}, {Fabrycky}, {Ford}, {Borucki},
  {Fressin}, {Marcy}, {Orosz}, {Rowe}, {Torres}, {Welsh}, {Batalha}, {Bryson},
  {Buchhave}, {Caldwell}, {Carter}, {Charbonneau}, {Christiansen}, {Cochran},
  {Desert}, {Dunham}, {Fanelli}, {Fortney}, {Gautier}, {Geary}, {Gilliland},
  {Haas}, {Hall}, {Holman}, {Koch}, {Latham}, {Lopez}, {McCauliff}, {Miller},
  {Morehead}, {Quintana}, {Ragozzine}, {Sasselov}, {Short}, \&
  {Steffen}}]{liss11a}
{Lissauer}, J.~J., {et~al.} 2011, \nat, 470, 53

\bibitem[{{Lissauer} {et~al.}(2013){Lissauer}, {Jontof-Hutter}, {Rowe},
  {Fabrycky}, {Lopez}, {Agol}, {Marcy}, {Deck}, {Fischer}, {Fortney}, {Howell},
  {Isaacson}, {Jenkins}, {Kolbl}, {Sasselov}, {Short}, \& {Welsh}}]{liss13}
---. 2013, \apj, 770, 131

\bibitem[{{Lithwick} {et~al.}(2012){Lithwick}, {Xie}, \& {Wu}}]{lith12}
{Lithwick}, Y., {Xie}, J., \& {Wu}, Y. 2012, \apj, 761, 122

\bibitem[{{Lopez} \& {Fortney}(2013)}]{lop13}
{Lopez}, E.~D., \& {Fortney}, J.~J. 2013, \apj, 776, 2

\bibitem[{{Mandel} \& {Agol}(2002)}]{man02}
{Mandel}, K., \& {Agol}, E. 2002, \apjl, 580, L171

\bibitem[{{Mazeh} {et~al.}(2013){Mazeh}, {Nachmani}, {Holczer}, {Fabrycky},
  {Ford}, {Sanchis-Ojeda}, {Sokol}, {Rowe}, {Zucker}, {Agol}, {Carter},
  {Lissauer}, {Quintana}, {Ragozzine}, {Steffen}, \& {Welsh}}]{maz13}
{Mazeh}, T., {et~al.} 2013, \apjs, 208, 16

\bibitem[{{Rauch} \& {Hamilton}(2002)}]{rh02}
{Rauch}, K.~P., \& {Hamilton}, D.~P. 2002, DAAS, 33, 938

\bibitem[{{Relles} (2013)}]{rel13}
{Relles}, http://exoplanet-science/com/KOI-152.html

\bibitem[{{Rogers} {et~al.}(2011){Rogers}, {Bodenheimer}, {Lissauer}, \&
  {Seager}}]{rog11}
{Rogers}, L.~A., {Bodenheimer}, P., {Lissauer}, J.~J., \& {Seager}, S. 2011,
  \apj, 738, 59

\bibitem[Rowe et al.(2013)]{row13} Rowe, J.F., et al. 2013, ApJ, submitted 

\bibitem[{{Sanchis-Ojeda} {et~al.}(2012){Sanchis-Ojeda}, {Fabrycky}, {Winn},
  {Barclay}, {Clarke}, {Ford}, {Fortney}, {Geary}, {Holman}, {Howard},
  {Jenkins}, {Koch}, {Lissauer}, {Marcy}, {Mullally}, {Ragozzine}, {Seader},
  {Still}, \& {Thompson}}]{san12}
{Sanchis-Ojeda}, R., {et~al.} 2012, \nat, 487, 449


\bibitem[{{Seager} \& {Mall{\'e}n-Ornelas}(2003)}]{sea03}
{Seager}, S., \& {Mall{\'e}n-Ornelas}, G. 2003, \apj, 585, 1038


\bibitem[{{Steffen} {et~al.}(2010){Steffen}, {Batalha}, {Borucki}, {Buchhave},
  {Caldwell}, {Cochran}, {Endl}, {Fabrycky}, {Fressin}, {Ford}, {Fortney},
  {Haas}, {Holman}, {Howell}, {Isaacson}, {Jenkins}, {Koch}, {Latham},
  {Lissauer}, {Moorhead}, {Morehead}, {Marcy}, {MacQueen}, {Quinn},
  {Ragozzine}, {Rowe}, {Sasselov}, {Seager}, {Torres}, \& {Welsh}}]{ste10}
{Steffen}, J.~H., {et~al.} 2010, \apj, 725, 1226

\bibitem[{{Steffen} {et~al.}(2012){Steffen}, {Ford}, {Rowe}, {Fabrycky},
  {Holman}, {Welsh}, {Batalha}, {Borucki}, {Bryson}, {Caldwell}, {Ciardi},
  {Jenkins}, {Kjeldsen}, {Koch}, {Pr{\v s}a}, {Sanderfer}, {Seader}, \&
  {Twicken}}]{ste12a}
{Steffen}, J.~H., {et~al.} 2012, \apj, 756, 186

\bibitem[{{Valencia} {et~al.}(2013){Valencia}, {Guillot}, {Parmentier}, \&
  {Freedman}}]{val13}
{Valencia}, D., {Guillot}, T., {Parmentier}, V., \& {Freedman}, R.~S. 2013,
  \apj, 775, 10


\bibitem[{{von Braun} {et~al.}(2011){von Braun}, {Boyajian}, {ten Brummelaar},
  {Kane}, {van Belle}, {Ciardi}, {Raymond}, {L{\'o}pez-Morales}, {McAlister},
  {Schaefer}, {Ridgway}, {Sturmann}, {Sturmann}, {White}, {Turner},
  {Farrington}, \& {Goldfinger}}]{von11}
{von Braun}, K., {et~al.} 2011, \apj, 740, 49

\bibitem[{{Wang} {et~al.}(2012){Wang}, {Ji}, \& {Zhou}}]{wan12}
{Wang}, S., {Ji}, J., \& {Zhou}, J.-L. 2012, \apj, 753, 170


\bibitem[{{Weiss} {et~al.}(2013){Weiss}, {Marcy}, {Rowe}, {Howard}, {Isaacson},
  {Fortney}, {Miller}, {Demory}, {Fischer}, {Adams}, {Dupree}, {Howell},
  {Kolbl}, {Johnson}, {Horch}, {Everett}, {Fabrycky}, \& {Seager}}]{weis13}
{Weiss}, L.~M., {et~al.} 2013, \apj, 768, 14

\bibitem[{{Winn}(2011)}]{winn11}
{Winn}, J.~N. 2011, in Exoplanets, ed. S.~{Seager} (University of Arizona
  Press), 55

\bibitem[{{Winn} {et~al.}(2011){Winn}, {Matthews}, {Dawson}, {Fabrycky},
  {Holman}, {Kallinger}, {Kuschnig}, {Sasselov}, {Dragomir}, {Guenther},
  {Moffat}, {Rowe}, {Rucinski}, \& {Weiss}}]{win11}
{Winn}, J.~N., {et~al.} 2011, \apjl, 737, L18

\bibitem[{{Wu} \& {Lithwick}(2013)}]{wu13}
{Wu}, Y., \& {Lithwick}, Y. 2013, \apj, 772, 74

\end{thebibliography}
\end{document}